 \documentclass[12pt]{article} 
\pdfoutput=1
\usepackage{graphicx}
\usepackage{fix-cm}%
\usepackage{smallsec}
\usepackage{mathptmx}
\usepackage{amssymb}%
\usepackage[footnotesize,labelfont=bf,labelsep=quad]{caption} 
\usepackage{amsmath}
\usepackage{hyperref}
\usepackage{theorem}  

{\theorembodyfont{\rmfamily}}
\sfcode`E=1000  
\newdimen\einr
\def\abs#1{\par\hangafter=1\hangindent=\einr\noindent
\hbox to\einr{\ignorespaces#1\hfill}\ignorespaces} 
\def\reals{{\mathbb R}} 
\def\0{{\bf 0}}
\def\1{{\bf 1}}
\def\T{^{\top}}

\def\tombstone{\hbox{\lower.4pt\vbox{\hrule\hbox{\vrule
  \kern7.6pt\vrule height7.6pt}\hrule}\kern.5pt}}
\def\MARK#1{\leavevmode\hbox to 3.5em{~~~\bf#1 :\hfill}\ignorespaces}
\def\3{\leavevmode\hbox to 2em{\hfill}\ignorespaces}
\def\4{\leavevmode\hbox to 3.5em{\hfill}\ignorespaces}
\newcount{\counter}
\def\5{\advance\counter by 1\noindent \leavevmode\hbox to
    1.5em{\hfill\scriptsize\the\counter}\hbox to
    2em{\hfill}\ignorespaces}
\def\6{\advance\counter by 1\noindent \leavevmode\hbox to
    1.5em{\hfill\lower.5ex\hbox{\small$^*$}\scriptsize~\the\counter}\hbox to
    2em{\hfill}\ignorespaces}
\def\7{{\hbox{*}}}%
\newdimen\emm\emm9.6pt 
\def\institute#1{}
\def\titlerunning#1{}
\begin{document}

\title{%
Game Theory Explorer -- Software for the Applied Game
Theorist 
} 


\author{Rahul Savani%
\thanks
{Department of Computer Science,
University of Liverpool,
Liverpool L69 3BX,
United Kingdom.
Email: rahul.savani@liverpool.ac.uk}
\and 
Bernhard von Stengel%
\thanks
{Department of 
Mathematics,
London School of Economics, London WC2A 2AE, United Kingdom.
Email: stengel@nash.lse.ac.uk}
}


\institute{R. Savani\at
Department of Computer Science,
University of Liverpool,
Liverpool L69 3BX,
United Kingdom.
Email: rahul.savani@liverpool.ac.uk
\and
B. von Stengel \at
Department of Mathematics,
London School of Economics, London WC2A 2AE, United Kingdom.
Email: stengel@nash.lse.ac.uk}


\date{\Large March 16, 2014}

\maketitle

\begin{abstract}
\noindent
This paper presents the ``Game Theory Explorer'' software
tool to create and analyze games as models of strategic
interaction.
A game in extensive or strategic form is created and nicely
displayed with a graphical user interface in a web browser.
State-of-the-art algorithms then compute all Nash equilibria
of the game after a mouseclick.
In tutorial fashion, we present how the program is used,
and the ideas behind its main algorithms.
We report on experiences with the architecture of the
software and its development as an open-source project.

\strut


%
%
\end{abstract}

\strut 

\section{Introduction}

Game theory provides mathematical concepts and tools for
modeling and analyzing interactive scenarios.  
In {\em noncooperative} game theory, the possible actions of the
players are represented explicitly, together with payoffs
that the players want to maximize for themselves.
Basic models are the {\em extensive} form represented by a
game tree with possible imperfect information represented by
information sets, and the {\em strategic} (or ``normal'')
form that lists the players' strategies, which they choose
independently, together with a table of the players' payoffs
for each strategy profile.

The central concept for noncooperative games is the {\em
Nash equilibrium} which prescribes a strategy for each
player that is optimal when the other players keep their
prescribed strategies fixed.
Every finite game has an equilibrium when players are
allowed to {\em mix} (randomize) their actions (Nash, 1951).
A game may have more than one Nash equilibrium.
Finding one or all Nash equilibria of a game is often
a laborious task.

In this paper, we describe the {\em Game Theory Explorer}
(GTE) software that allows to create games in extensive or
strategic form, and to compute their Nash equilibria.
GTE is primarily intended for the applied game
theorist who is not an expert in equilibrium computation,
for example an experimental economist who designs
experiments to test if subjects play equilibrium strategies.
The user can easily vary the parameters of the
game-theoretic model, and study more complex games, because
their equilibrium analysis is quickly provided by the
algorithm.
The analysis of a game with general mathematical parameters
is also aided by knowing its equilibria for specific
numerical values of those parameters.
As our exposition will demonstrate, GTE can also be used for
more theoretical research in game theory, for example on
strategic stability (see Section~\ref{s-strategic}).
The ease of creating, displaying, and analyzing games
suggests GTE also as an {\em educational tool} for game
theory.

Computing equilibria is a main research topic of the
authors, and in later sections we will explain some of our
results on finding equilibria of two-player games in
strategic form (Avis et al., 2010) and extensive form
based on the ``sequence form'' (von Stengel, 1996).
Scientific algorithms are often implemented as prototopes to
show that they work, as done, for example, by Audet et al.\
(2001), or von Stengel, van den Elzen, and Talman (2002).
However, providing a robust user interface to create games
is much more involved, and necessary to make such algorithms
useful for a wider research community.
In particular, the drawing of game trees should be done with
a friendly {\em graphical user interface} (GUI) where the
game tree can be created and seen on the screen, and be
stored, retrieved, and changed in an intuitive manner.
This is one of the purposes of the GTE software presented in
this article.

An existing suite of software for game-theoretic analysis is
{\em Gambit} (McKelvey, McLennan, and Turocy, 2010).
Gambit has been developed over the course of nearly 25 years
and presents a library of solution algorithms, formats for
storing games, ways to program the creation of games with
the help of the Python programming language, and a GUI for
creating game trees.
It is open-source software that is free to use and that can
be extended by anyone.
Given the mature state of Gambit and the joint research
interests and close contacts with its developers, it is
clear that any improvements offered by GTE should eventually
be integrated into Gambit.

Other existing game solvers are 
GamePlan (Langlois, 2006) and 
XGame (Belhaiza, Mve, and Audet, 2010).

The main difference of GTE to Gambit is the provided access
to the software and the user interface.  
In terms of access, Gambit needs to be downloaded and
installed; it is offered on the main personal computing
platforms Windows, Linux or Mac.
Getting the program to run may require some patience and
technical experience with software installions, which may
present a ``barrier to entry'' for its use.
In contrast, GTE is started in a {\em web browser}
via the web address
\url{http://www.gametheoryexplorer.org}.
All interaction with the software is via the browser
interface.
The created games and their output can be saved as files by
the user on their local computer.
This avoids the technical hurdles of installing software on
the user side, and simplifies updating the software.

The graphical display of game trees in GTE is user-friendly
and can be customized, such as growing the tree in the
vertical or horizontal direction.
GTE can even be used just as a drawing tool for games, which
can be exported as pictures to file formats for use in
papers or presentations.

Providing application software via the web has the following
disadvantages compared to installed software.
First, a higher complexity of the program for the required
communication over the internet; however, manifold standard
solutions are freely available.
Second, limited control over the user's local computing
resources for security reasons.
This is an issue because equilibrium computation for larger
games is computationally intensive.
For that reason, this computation takes place
on the central public server
rather than the user's client computer; we will explain
the technical issues in Section~\ref{s-software}.
Third, in our case, currently very limited use of the
existing Gambit software.
We envisage a loosely coupled integration, in particular
with Gambit's game solvers.
GTE is very much under development in this respect.

We describe GTE, first from the perspective of the user,
with an example in Section~\ref{s-example}, and the general
creation of extensive and strategic-form games in Sections
\ref{s-extensive} and~\ref{s-strategic}.
We explain the main algorithms for finding all equilibria
for games in strategic form in Section~\ref{s-algostrat},
including issues for handling larger games where one has to
restrict oneself to finding sample equilibria (which, in
particular, does not allow to decide if the game has a
unique equilibrium).
For extensive games, the computation of behavior strategies,
which are exponentially less complex than mixed strategies,
is outlined in Section~\ref{s-sequenceform}.
The software architecture and the communication between
server and client computers is discussed, with as little
technical jargon as possible, in Section~\ref{s-software}.
In conclusion, we mention the difficulties of incentives and
funding for implementing user-friendly scientific software,
and call for contributions of volunteers.

\section{Example of using GTE}
\label{s-example}

In this section we describe a simple $2\times 2$ game, due
to Bagwell (1995), together with its ``commitment'' or
``Stackelberg'' variant which has more strategies.
This game is created and analyzed very simply with GTE,
where we demonstrate the computed equilibria.
Our detailed analysis may also serve as a tutorial
introduction to the basics of noncooperative game theory.
For a textbook on game theory see Osborne (2004).

A basic model of a noncooperative game is a game in {\em
strategic form}.
Each player has a finite list of strategies.
Players choose simultaneously a strategy each, which defines
a strategy {\em profile}, with a given {\em payoff} to each
player.
The strategic form is the table of payoffs for all strategy
profiles.
For two players, the strategies of player~1 are the $m$ rows and
those of player~2 the $n$ columns of a table, which in each cell
has a payoff pair.
The two $m\times n$ matrices of payoffs to player 1 and~2
define such a two-player game in strategic form, which is
also called a bimatrix game.

\begin{figure}[hbt]
\begin{minipage}{.38\hsize}
\includegraphics[height=11\emm]{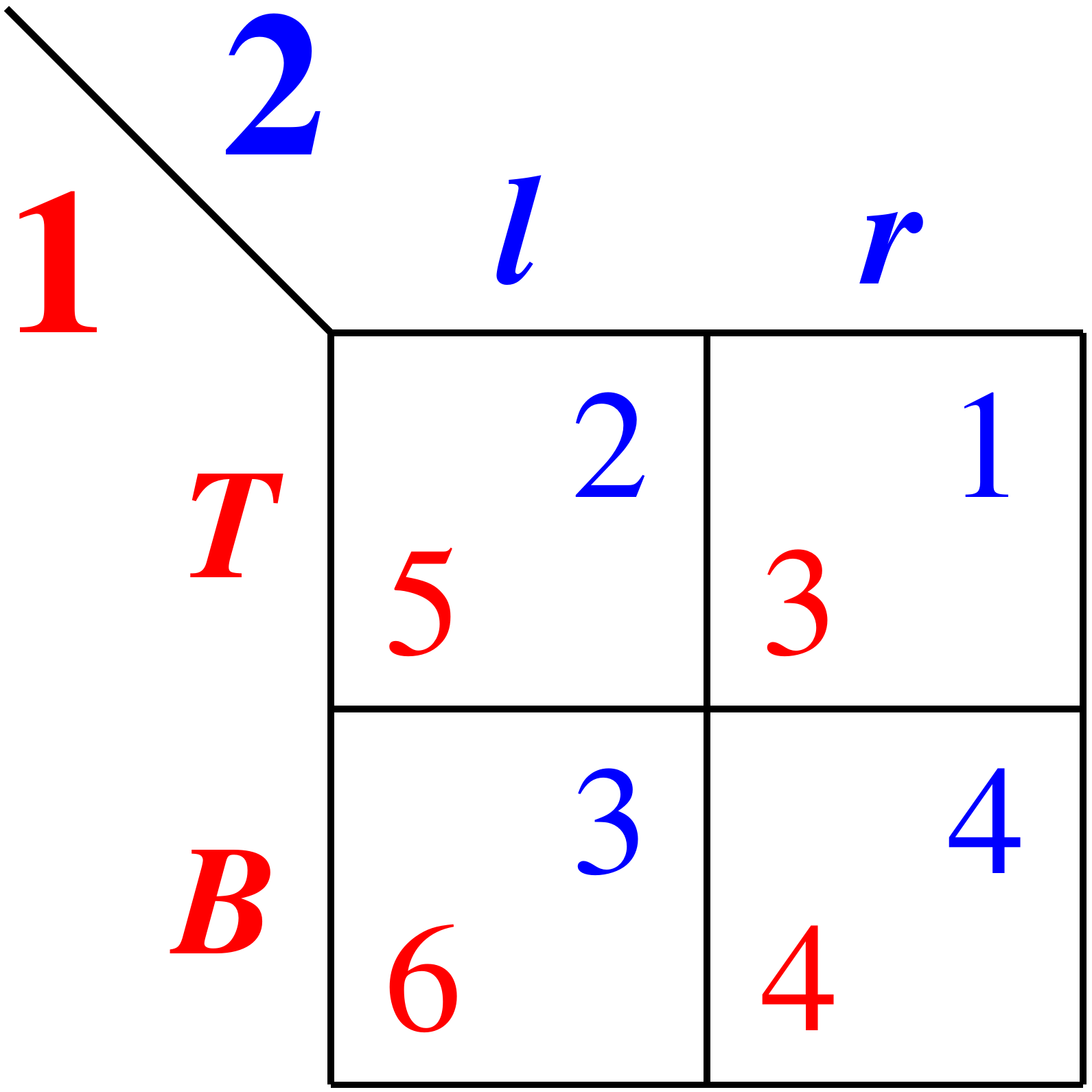}
\vskip 8ex
\includegraphics[height=11\emm]{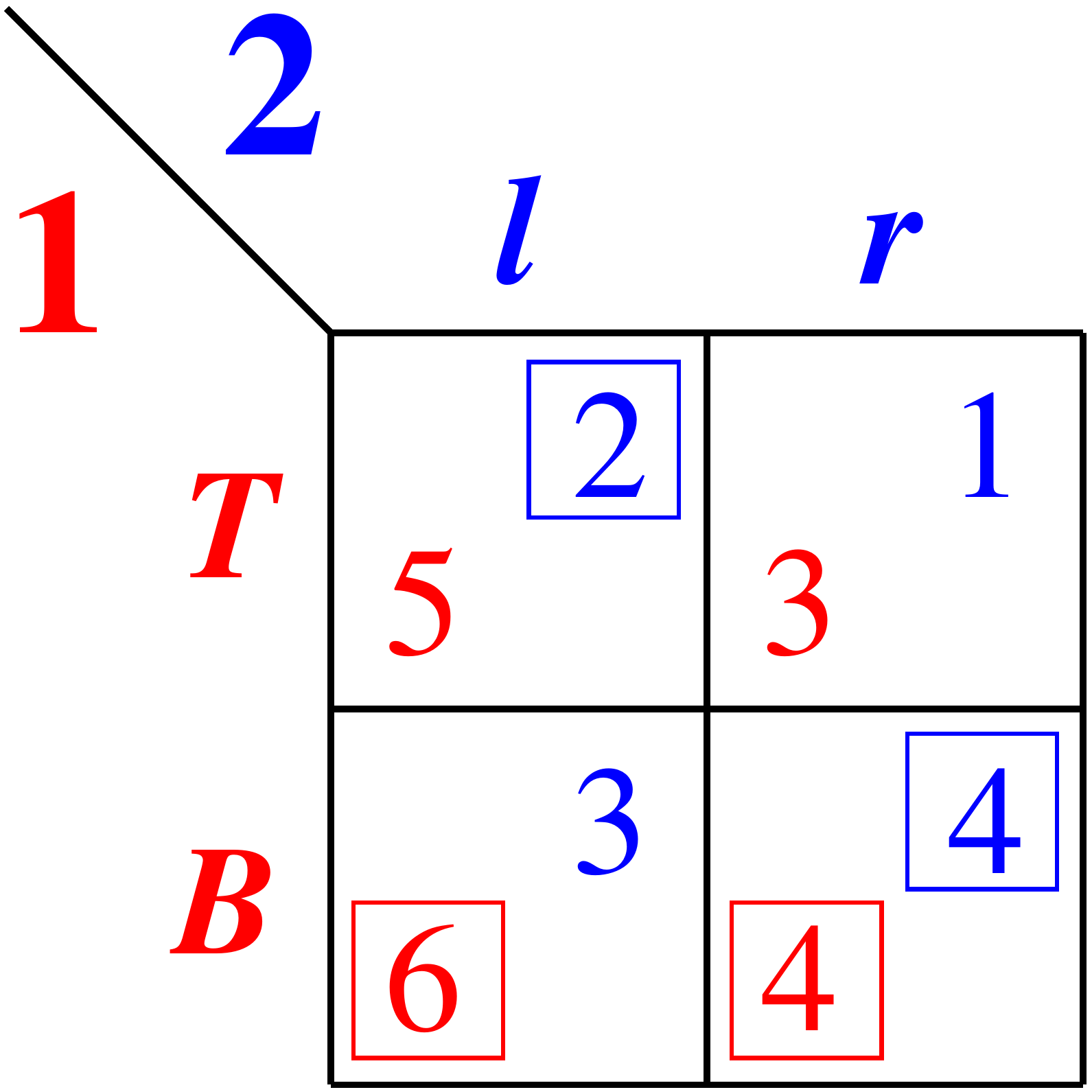}
\end{minipage}%
\begin{minipage}{.62\hsize}
\footnotesize
\begin{verbatim}
Strategic form: 
2 x 2 Payoff player 1

  l r
T 5 3
B 6 4

2 x 2 Payoff player 2

  l r
T 2 1
B 3 4

EE = Extreme Equilibrium, EP = Expected Payoffs

Rational:
EE 1 P1: (1) 0 1 EP= 4 P2: (1) 0 1 EP= 4 

Decimal:
EE 1 P1: (1) 0 1.0 EP= 4.0 P2: (1) 0 1.0 EP= 4.0 

Connected component 1:
{1}  x  {1}
\end{verbatim}
\end{minipage} 
\caption{Top left: Graphical output of a $2\times2$ bimatrix game.
Bottom left: Best response payoffs surrounded by boxes
(currently not part of GTE output).
Right: Output of the computed equilibria of this game.}
\label{f1}
\end{figure}

Fig.~\ref{f1} shows a $2\times 2$ game with the payoff to
player~1 shown in the bottom left corner and the payoff to
player~2 in the top right corner of each cell.
In GTE, such a game is entered by giving the two payoff
matrices (Fig.~\ref{fstratinput} shows the input screen
for a larger example), with a graphical display as shown at
the top left of Fig.~\ref{f1}.
At the bottom left the same table is shown with a box around
each payoff that is maximal against the respective strategy
of the other player (these boxes are currently not part of
GTE output).
This shows that the bottom strategy $B$ of player~1 is the
only {\em best response} (payoff-maximizing strategy)
against both columns $l$ and $r$ of player~2.
For player~2, the best responses are $l$ against $T$ and
$r$ against~$B$.

In this game, strategy $T$ is strictly dominated by $B$ and
can therefore be disregarded because it will never be played
by player~1.
The best response $r$ against $B$ then defines the strategy
profile $(B,r)$ as the unique Nash equilibrium of the game,
that is, a pair of mutual best responses.
The text output at the right of Fig.~\ref{f1}, which pops
up in a window after clicking a button that starts the
equilibrium computation, shows the two payoff matrices and
the Nash equilibria.
The equilibrium strategies are shown as vectors of
probabilities, both in rational output as exact fractions of
integers and in decimal.
In this case there is only one equilibrium, which is the
pair of vectors $(0,1)$ and $(0,1)$ that 
describe the probabilities that player~1 plays his pure
strategies $T$ and $B$ and player~2 her strategies $l$
and~$r$, respectively.
This pair of strategies also defines the unique equilibrium
component shown as \verb.{1} x {1}., an output which is of
more interest in the example of Fig.~\ref{f3} below.

\begin{figure}[hbt]
\strut\hfill
\includegraphics[height=17\emm]{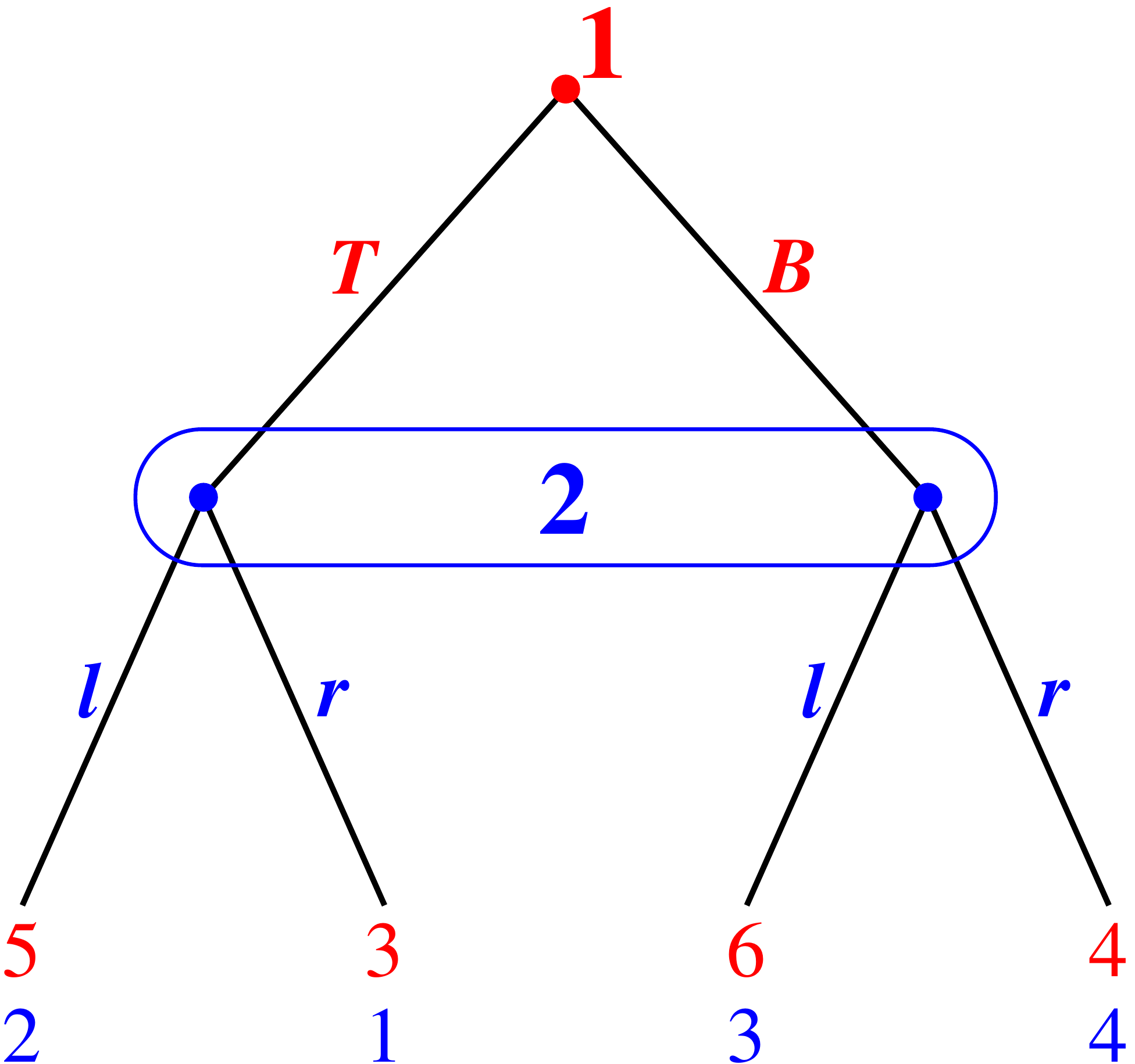}
\hfill\strut
\caption{The game in Fig.~\ref{f1} as an extensive game
with an information set for player~2 who is uninformed about
the move of player~1.}
\label{f2}
\end{figure}

The game in Fig.~\ref{f1} can also be represented in
{\em extensive form}, as shown in Fig.~\ref{f2} which is
automatically generated by GTE.
An extensive game is a game tree with nodes as game states
and outgoing edges as moves chosen by the player to move at
that node.
{\em Information sets}, due to Kuhn (1953), describe a
player's lack of information about the current game state
and have the same outgoing moves at each node.
Here, player~2 is not informed about the move of player~1,
in accordance with the players' simultaneous choice of moves
in the strategic form.

\begin{figure}[hbt]
\includegraphics[height=17\emm]{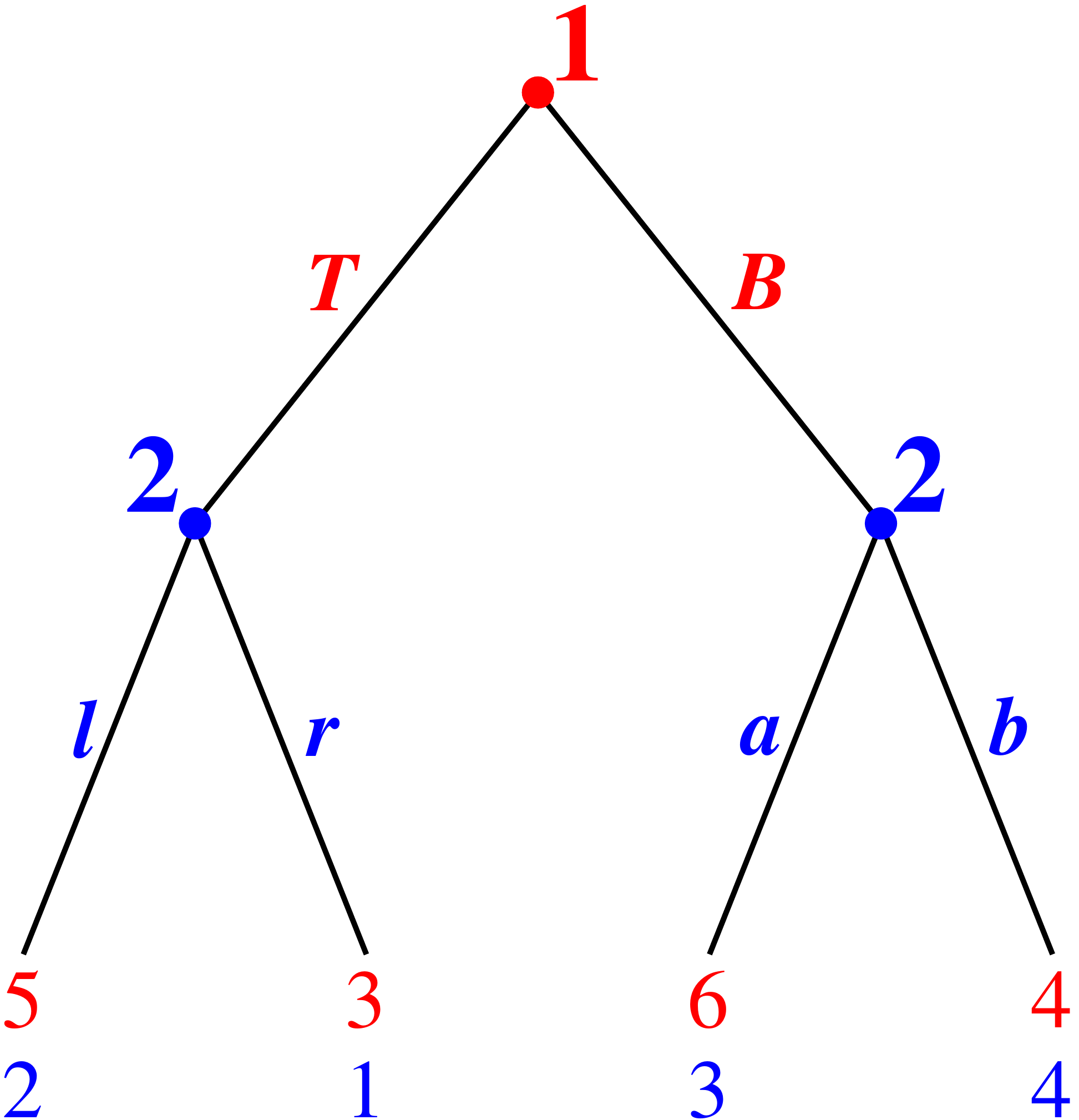} 
\hfill
\raise3\emm\hbox{
\includegraphics[height=11\emm]{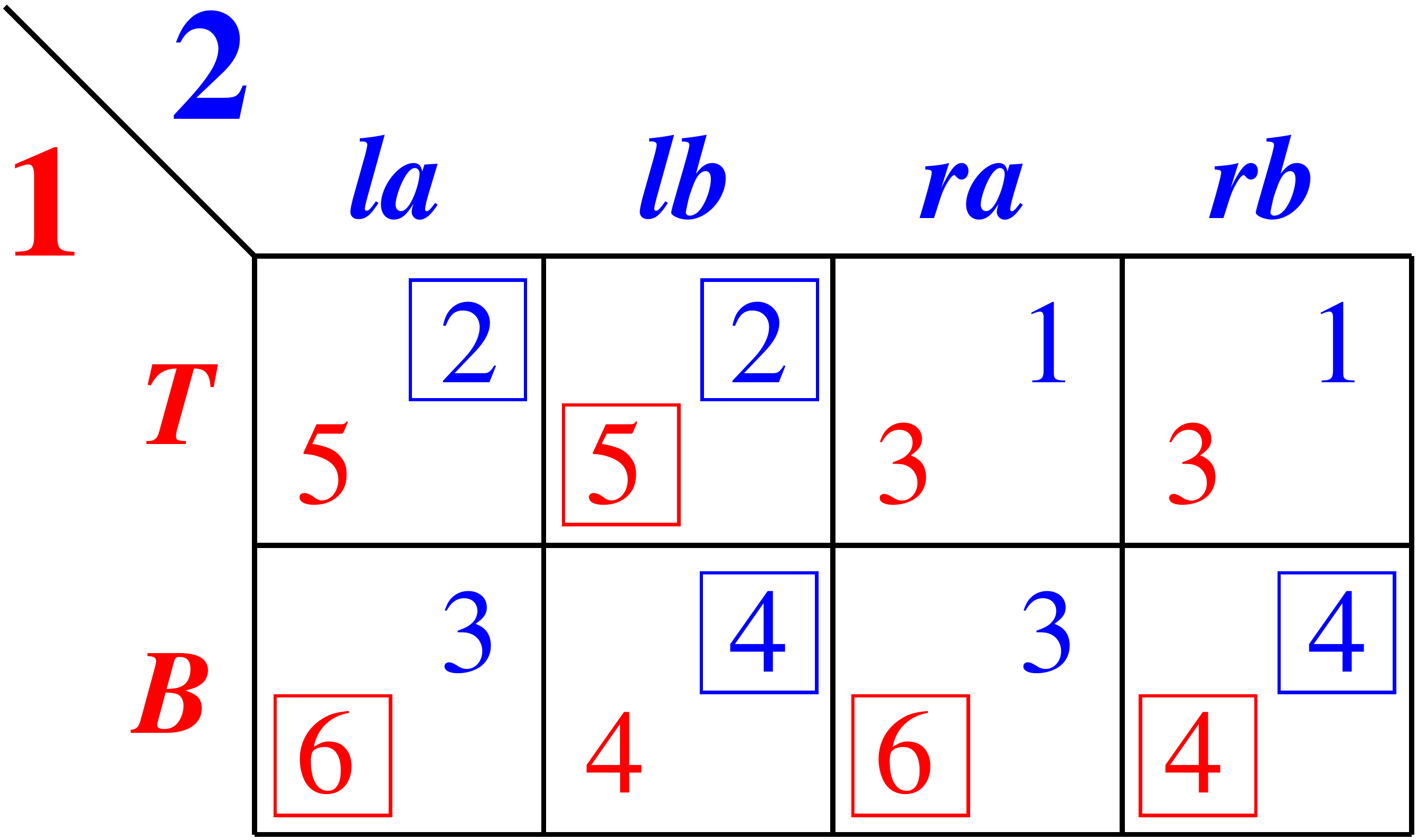}}
\caption{Left: Commitment version game of the game in
Fig.s \ref{f1} and~\ref{f2} where player~2 is informed
about the first move of player~1.
Right: strategic form of this game as generated by GTE
(except for the boxes around the best-response payoffs).}
\label{f3}
\end{figure}

The game in Fig.~\ref{f1} is an example due to Bagwell
(1995). 
It is a simple version of a ``Cournot'' game.
In the corresponding ``Stackelberg'' or {\em commitment}
game, player~1 is a leader who commits to his strategy,
about which player~2, as a follower, is informed.
The game tree, and thus the game, is changed by becoming a
game of perfect information where each information set is a
singleton.\footnote{GTE does not show singleton information
sets as ovals that contain a single node, only information
sets with two or more nodes.} 
To change the game in this way the information set is
dissolved, which is a simple operation in GTE.
The new game tree is shown on the left in Fig.~\ref{f3}.
Then the moves of player~2, who can react to the choice of
player~1, get new names, here at the right node $a$ and $b$
instead of the original moves $l$ and $r$ at the original
information set that remain the moves at the left node of
player~2.

A game tree with perfect information can be solved by 
``backward induction'' which defines a {\em subgame perfect
equilibrium} or SPE, which is indeed a Nash equilibrium.
Here, player~2's optimal moves are $l$ and $b$, which
defines her strategy $lb$.
Given these choices of player~2, the optimal move of player~1
is~$T$.
This defines the SPE $(T,lb)$.
In general, a strategy in an extensive game specifies a move
at each information set of the player, so player~2 has the
four strategies $la$, $lb$, $ra$, $rb$ listed as columns in
the strategic form on the right in Fig.~\ref{f3}, which is
generated by GTE.
Because player~1 has only one information set given by the
singleton that contains the root node of the tree, his
strategies are just the moves $T$ and $L$.
The SPE $(T,lb)$ is one of the cells in the strategic
form with the two best-response payoffs $5$ and $2$ for the
two players.

In a game tree with perfect information and different
payoffs at each terminal node, backward induction defines a
unique SPE.
However, the game has in general additional Nash equilibria
that are not subgame perfect.
In this example, the strategy pair $(B,rb)$ is also an
equilibrium, which can be seen from the strategic form.
Here player~2 chooses the right move $r$ and $b$ at both
information sets, and the best response of player~1 is
then~$B$, with payoffs $4$ and $4$ to the two players.
This is an equilibrium because neither player can
unilaterally improve his or her payoff, under the crucial
assumption of equilibrium that the strategy of the other
player stays fixed:
When player~1 chooses $T$ instead of $B$ and player~2 plays
$rb$, then player~1 receives a payoff of $3$ rather than
$4$ and therefore prefers to stay with~$B$.
In turn, $b$ is an optimal move when player~1 chooses~$B$.
Player~2 cannot improve her payoff by changing from $r$
to~$l$ because that part of the game tree is not reached due
to the move $B$ by player~1.
This equilibrium is in effect the equilibrium $(B,r)$ in
the original simultaneous game in Fig.~\ref{f1} translated
to the commitment game where player~2, even though she can
now react to the move of player~1, always chooses the
equivalent of the original move~$r$.
However, this equilibrium is not subgame perfect because it
prescribes the suboptimal move $r$ in the subgame that
starts with the left node of player~2. 

In addition to these two pure-strategy equilibria, the game
in Fig.~\ref{f3} has additional equilibria where player~2
makes a random choice at the node that is unreached due to
the move of player~1.
Because the node is unreached, any choice of player~2 is
optimal because it has no effect on her payoffs, but that
random choice must not change the preference of player~1 for
his move in order to keep the equilibrium property.
Move $B$ of player~1, followed necessarily by move $b$ of
player~2, is optimal for player~1 as long as his expected
payoff for $T$ is at most~$4$, which is the case whenever
player~2 makes move $r$ with probability at least $1/2$.
The two extreme cases are the pure strategy equilibrium
$(B,rb)$ already discussed and the mixed strategy
equilibrium where player~1 chooses $B$ and player~2 mixes
between $lb$ and $rb$ with probability $1/2$ each.
This is represented by the probability vector
$(0,1/2,0,1/2)$ for the four strategies of player~2.
Similarly, an equilibrium that has the same outcome with
payoffs $5$ and $2$ as the SPE but a suboptimal random choice 
between $a$ and $b$ of player~2 is $(T, (1/2,1/2,0,0))$
where player mixes between $la$ and $lb$ with
probability $1/2$ each; $1/2$ is the largest probability
that player~2 can assign to $a$ so that $T$ stays a best
response of player~1.

The preceding analysis is straightforward and simple, but a
complete list of all equilibria is nevertheless of interest.
This is provided by GTE with the following output:

{\footnotesize
\begin{verbatim}
    Strategic form: 
    2 x 4 Payoff player 1

      la lb ra rb
    T  5  5  3  3
    B  6  4  6  4

    2 x 4 Payoff player 2

      la lb ra rb
    T  2  2  1  1
    B  3  4  3  4

    EE = Extreme Equilibrium, EP = Expected Payoffs

    Rational:
    EE 1 P1: (1) 0 1 EP= 4 P2: (1)   0 1/2 0 1/2 EP= 4 
    EE 2 P1: (1) 0 1 EP= 4 P2: (2)   0   0 0   1 EP= 4 
    EE 3 P1: (2) 1 0 EP= 5 P2: (3)   0   1 0   0 EP= 2 
    EE 4 P1: (2) 1 0 EP= 5 P2: (4) 1/2 1/2 0   0 EP= 2 

    Decimal:
    EE 1 P1: (1)   0 1.0 EP= 4.0 P2: (1)   0 0.5 0 0.5 EP= 4.0 
    EE 2 P1: (1)   0 1.0 EP= 4.0 P2: (2)   0   0 0 1.0 EP= 4.0 
    EE 3 P1: (2) 1.0   0 EP= 5.0 P2: (3)   0 1.0 0   0 EP= 2.0 
    EE 4 P1: (2) 1.0   0 EP= 5.0 P2: (4) 0.5 0.5 0   0 EP= 2.0 

    Connected component 1:
    {1}  x  {1, 2}

    Connected component 2:
    {2}  x  {3, 4} 
\end{verbatim}}

This output gives the four ``extreme'' equilibria described
above, in both rational and decimal description.
Each equilibrium strategy of a player is preceded by an
identifying number in parentheses such as 
\verb.(1). and \verb.(2). for the two strategies of
player~1, each of which appears in two equilibria.
All four equilibrium strategies of player~2 are distinct,
marked with \verb.(1). to \verb.(4)..
The connected components listed at the end of the output
show how these extreme equilibria can be arbitrarily
combined:
The first connected component
\verb. {1}  x  {1, 2} .
says that strategy \verb.(1). of player~1, which is $(0,1)$
(the pure strategy $B$), together with {\em any convex
combination} of strategies \verb.(1). and \verb.(2). of
player~2, which are the strategies 
$(0,1/2,0,1/2)$ and $(0,0,0,1)$ (the latter being the pure
strategy $rb$), defines an equilibrium.
This and the other connected component
\verb. {2}  x  {3, 4} .
describe the full set of Nash equilibria of the game.

In principle, GTE provides such a complete description of
all equilibria for any two-player game, except that the
computation time, which is in general exponential in the
size of the game, may be prohibitively long for larger
games.

\section{Creating and analyzing extensive form games}
\label{s-extensive}

In this section, we describe the construction in GTE of an
extensive game which corresponds to the game of the previous
section but where the first player's commitment is
imperfectly observed, as studied by Bagwell (2005).
The game is created in five stages:
Drawing the raw tree, assigning players, combining nodes
into information sets, defining moves and chance
probabilities, and setting payoffs.
Graphical and other more permanent settings, such as the
orientation of the game tree, can also be changed.
After explaining the GUI operations, we consider the
interesting equilibrium structure of the constructed game.

\begin{figure}[hbt]
\leavevmode
\raise4\emm\hbox{{\includegraphics[height=8\emm]{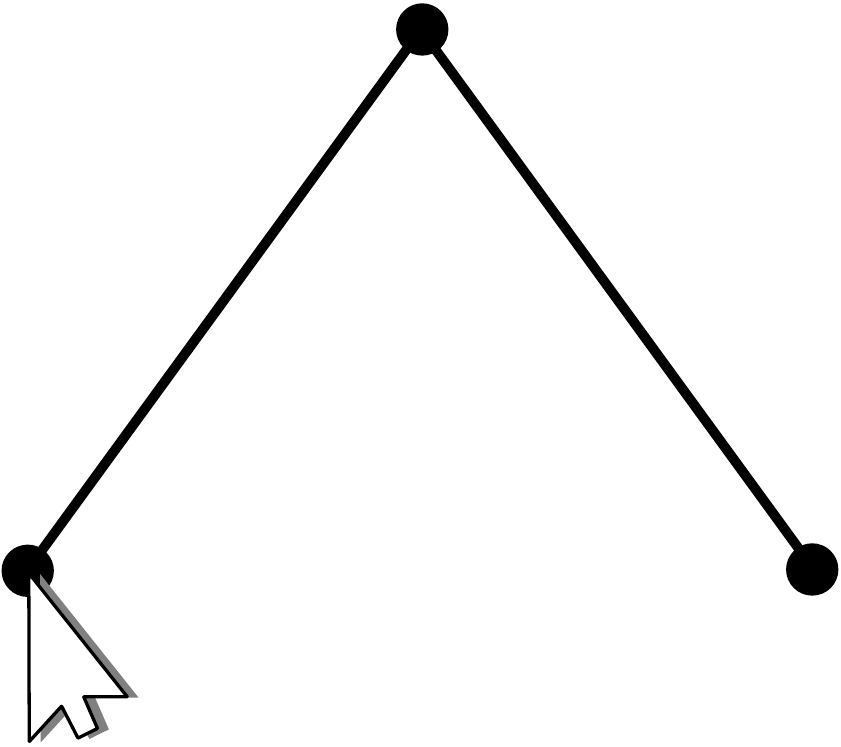}}}
\hfill
\includegraphics[height=12\emm]{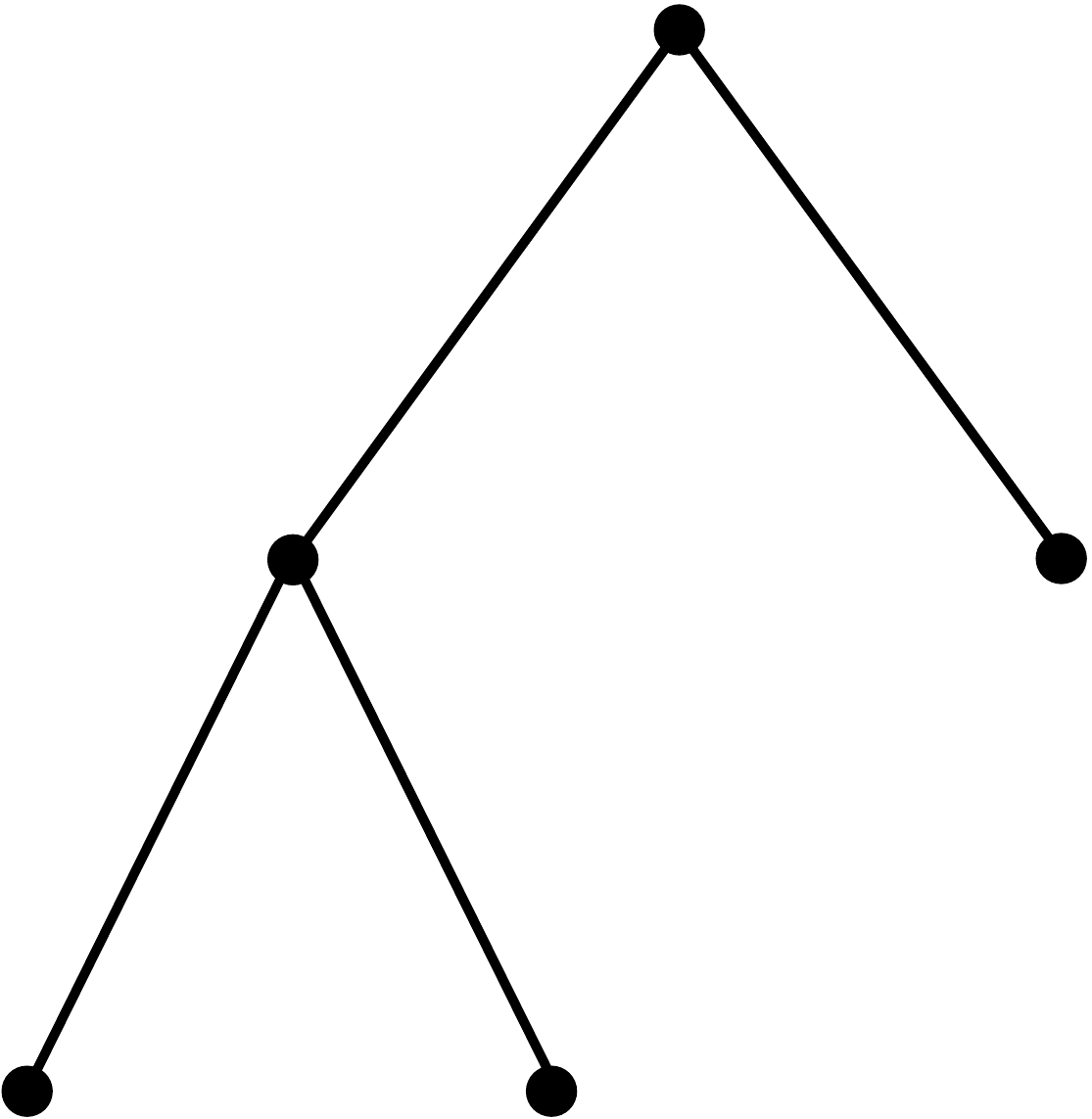}

\vskip-5ex
\strut\hfill
\includegraphics[width=23\emm]{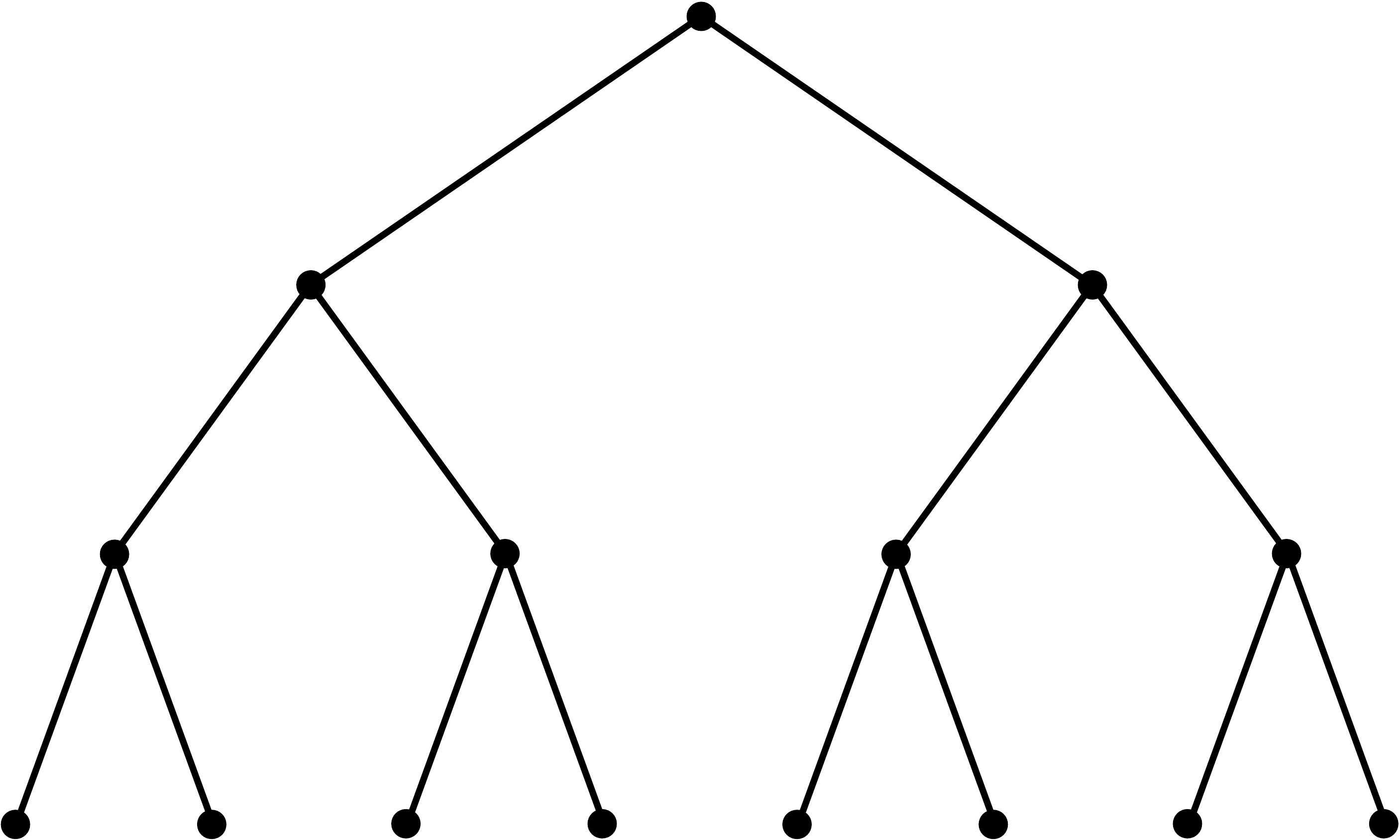}
\hfill\strut
\hfill\strut
\hfill\strut
\caption{Constructing a tree in stages by clicking on nodes,
as shown with the starting tree at the top left, second stage
at the top right, and final stage at the bottom.
In this tree, every nonterminal node has two children, but in
general it may have any number of children.}
\label{ftree}
\end{figure}

The first stage of creating a new extensive game defines the
tree structure, beginning from a simple tree that has a root
node with two children, as shown in Fig.~\ref{ftree}.
Clicking on a leaf (terminal node) creates two children for 
that node, and clicking on a nonterminal node creates an
additional child.

\begin{figure}[hbt]
\strut\hfill
\includegraphics[width=25\emm]{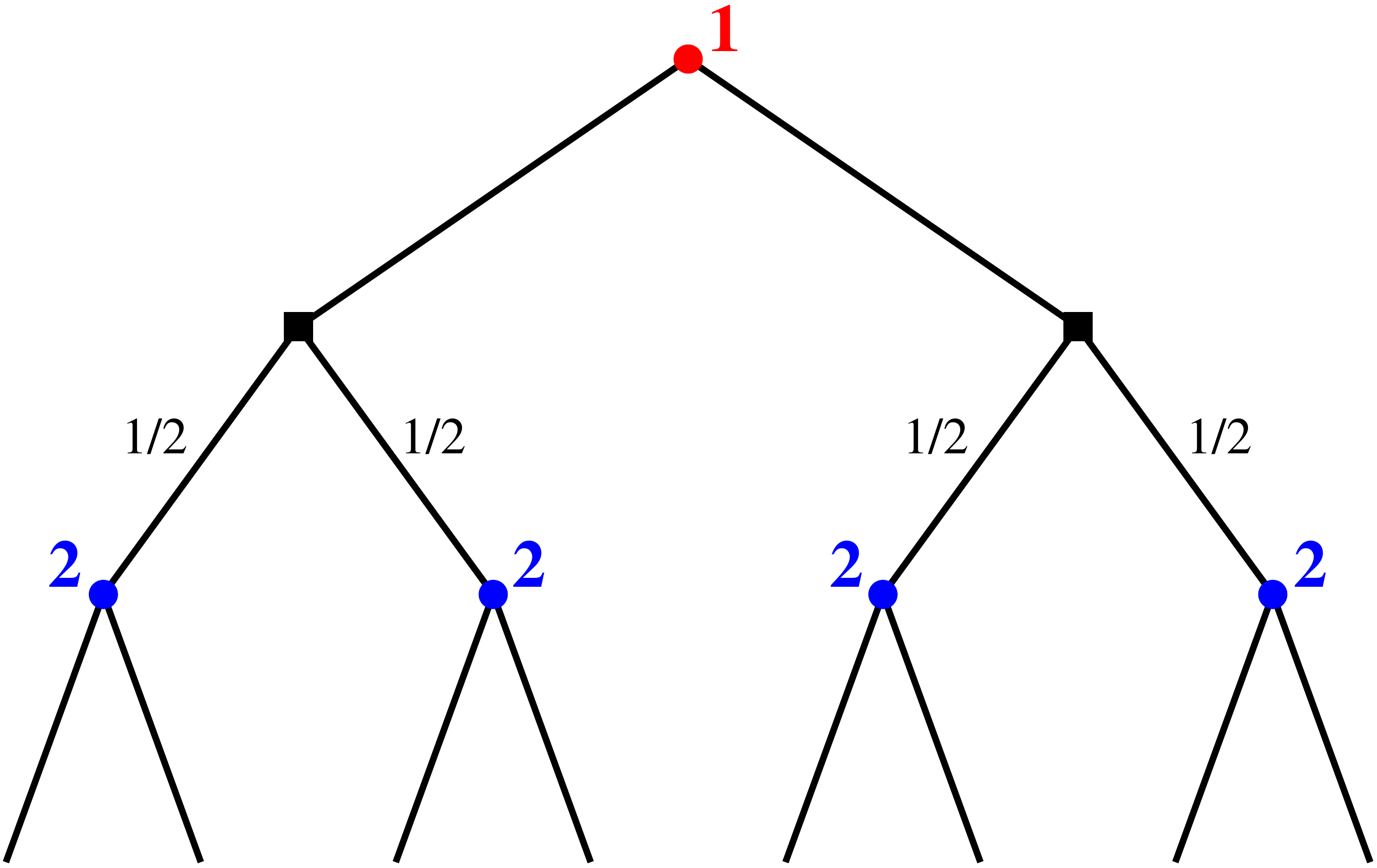}
\hfill\strut
\caption{Choosing the player to move for each node.
The square nodes are chance nodes, with initially uniform
probabilities for the outgoing edges.
}
\label{fplayers}
\end{figure}

The next stage is to select players, by selecting a ``player
assignment'' button for each player, where the name of the
player can also be changed, for example to ``Alice'' and
``Bob'' instead of the defaults ``1'' and ``2'' (which we
have not done).
Clicking on a nonterminal node then asigns the player
(originally unassigned with a black node), for example
player~2 for the four nonterminal nodes closest to the
leaves in Fig.~\ref{fplayers}.
Assigning a chance node, represented by a black square,
defines per default uniform probabilities on the outgoing
edges, which can be changed.  
All nonterminal nodes have to be assigned a player before
the next stage can be reached.
It is possible to go back to a previous stage at any time,
here to alter the tree structure by adding or deleting
nodes.

\begin{figure}[hbt]
\strut\hfill
\includegraphics[width=35\emm]{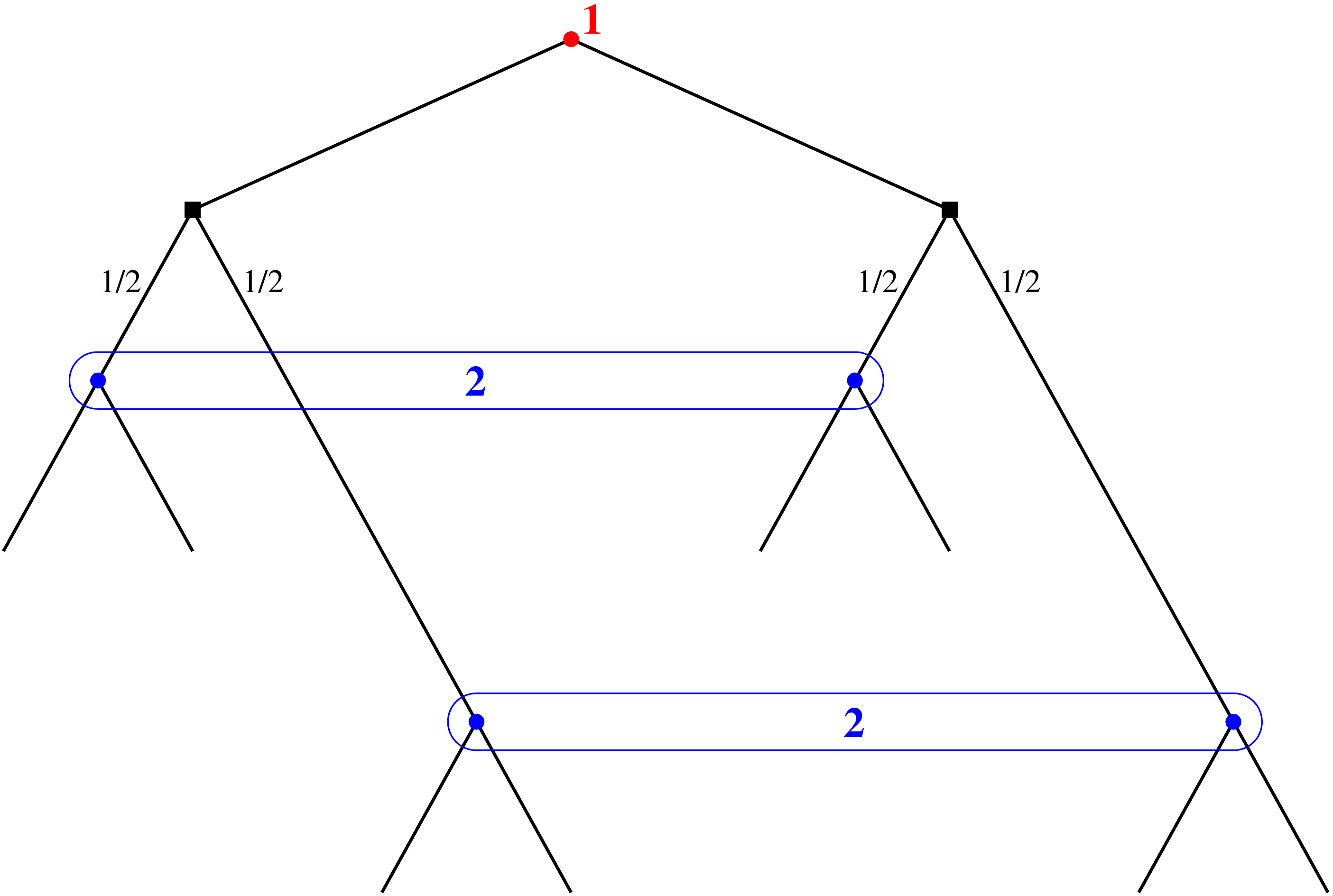}
\hfill\strut
\caption{Creating information sets, which are automatically
drawn so as to minimize the number of crossing edges.
}
\label{finfosets}
\end{figure}

\begin{figure}[hbt]
\includegraphics[width=\hsize]{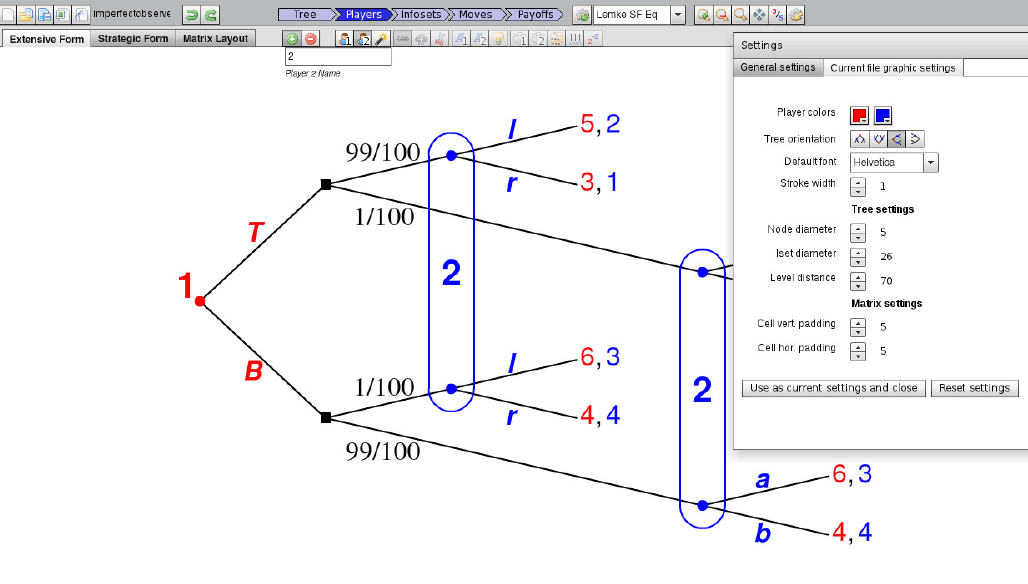} 
\caption{Changing the graphics settings, in this case to
a left-to-right tree orientation and a Helvetica font.
}
\label{fsettings}
\end{figure}

The next stage is to create information sets by clicking on
two nodes (or information sets) of the same player, which
are then merged into a single information set.
It is necessary that the respective nodes have the same
number of children because these will be the moves at the
newly created information set.  
When an information set is created, the program adjusts, as
far as possible, the levels of nodes in the tree so that all
nodes in the information set are at the same level and the
information set appears horizontally (if the game tree grows
in the vertical direction, otherwise vertically as in
Fig.~\ref{fsettings}).
In addition, crossings between edges and information sets
are minimized.
Fig.~\ref{finfosets} shows the resulting game tree, which
at this stage has its final shape, except for the definition
of moves and payoffs.

\begin{figure}[hbt]
\strut\hfill
\includegraphics[width=25\emm]{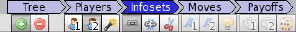} 
\hfill\strut
\caption{Browser headline bar that guides through the stages
of creating a game tree, here indicating the ``information
set'' creation stage.} 
\label{fheadline}
\end{figure}

Fig.~\ref{fheadline} shows a ``headline bar'' that
indicates the current stage of the creation of the game
tree, which defines either the tree, the players, the
information sets, the moves, or the payoffs.
The location of the headline bar relative to the entire
browser window can be seen in Fig.~\ref{fsettings}.
In Fig.~\ref{fheadline}, the current stage is that of
creating information sets.
Underneath the stage indicator are buttons that define the
{\em mode} of operation for the computer mouse.
The default mode at the ``information set'' stage is that of
merging two nodes (more generally the information set that
they are currently contained in) into a larger information
set.
The two other mode buttons at this stage are to dissolve an
information set into singletons, or (indicated by the
scissors) to cut it into two smaller sets.

The default mode for the earlier ``tree'' stage is to add
children to a node (indicated by the $\oplus$ sign).
The alternative mode button (with the $\ominus$ sign) is to
delete a node and all its descendants.
At the ``players'' stage, the mode buttons correspond to the
player (including chance) to be assigned to the node when
clicking on the node.  

\begin{figure}[hbt]
\strut\hfill
\includegraphics[width=40\emm]{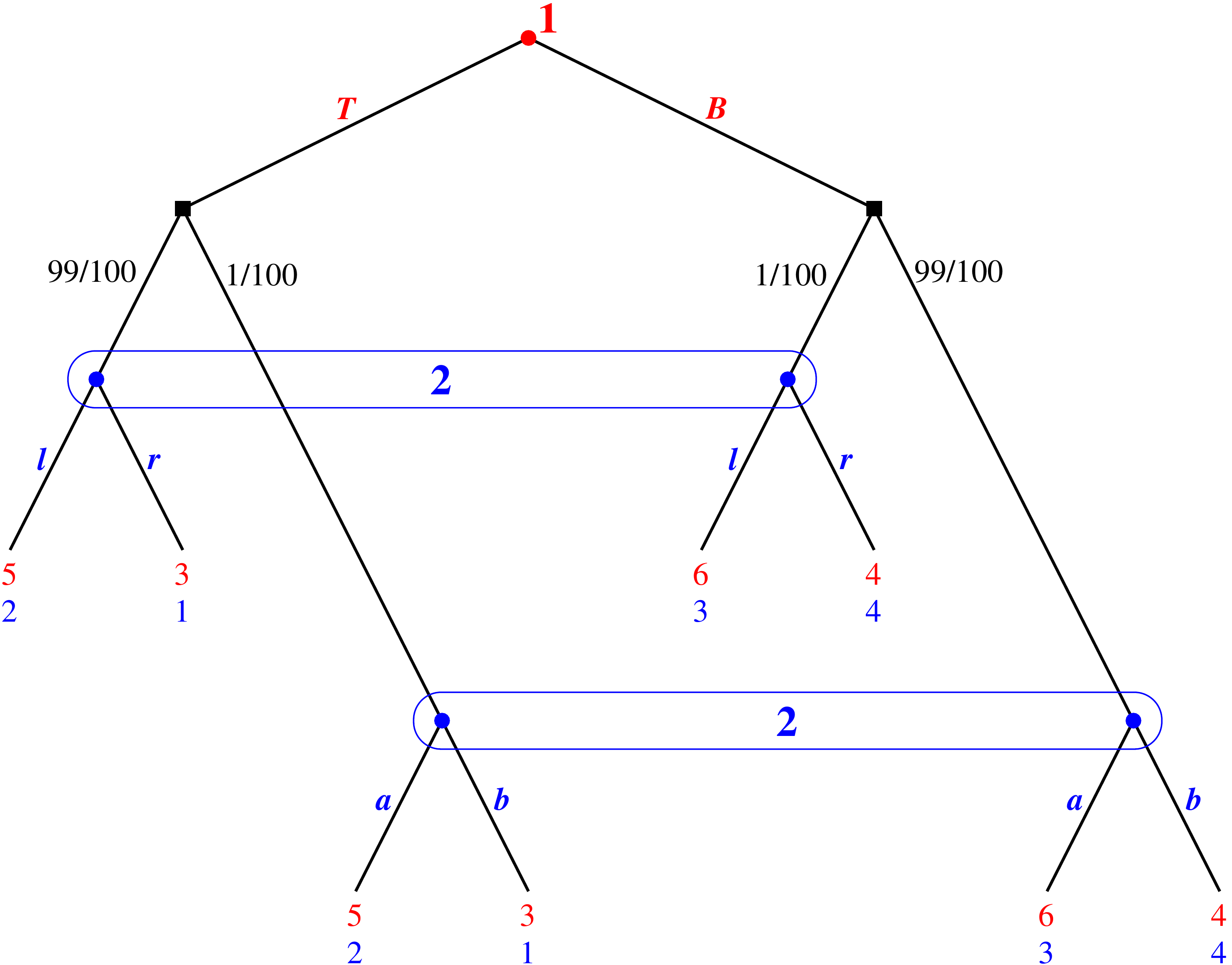}
\hfill\strut
\caption{The final game tree after adding moves, modified
chance probabilities, and payoffs.
}
\label{fpayoffs}
\end{figure}

\begin{figure}[hbt]
\noindent
\raise1\emm\hbox{\includegraphics[height=13\emm]{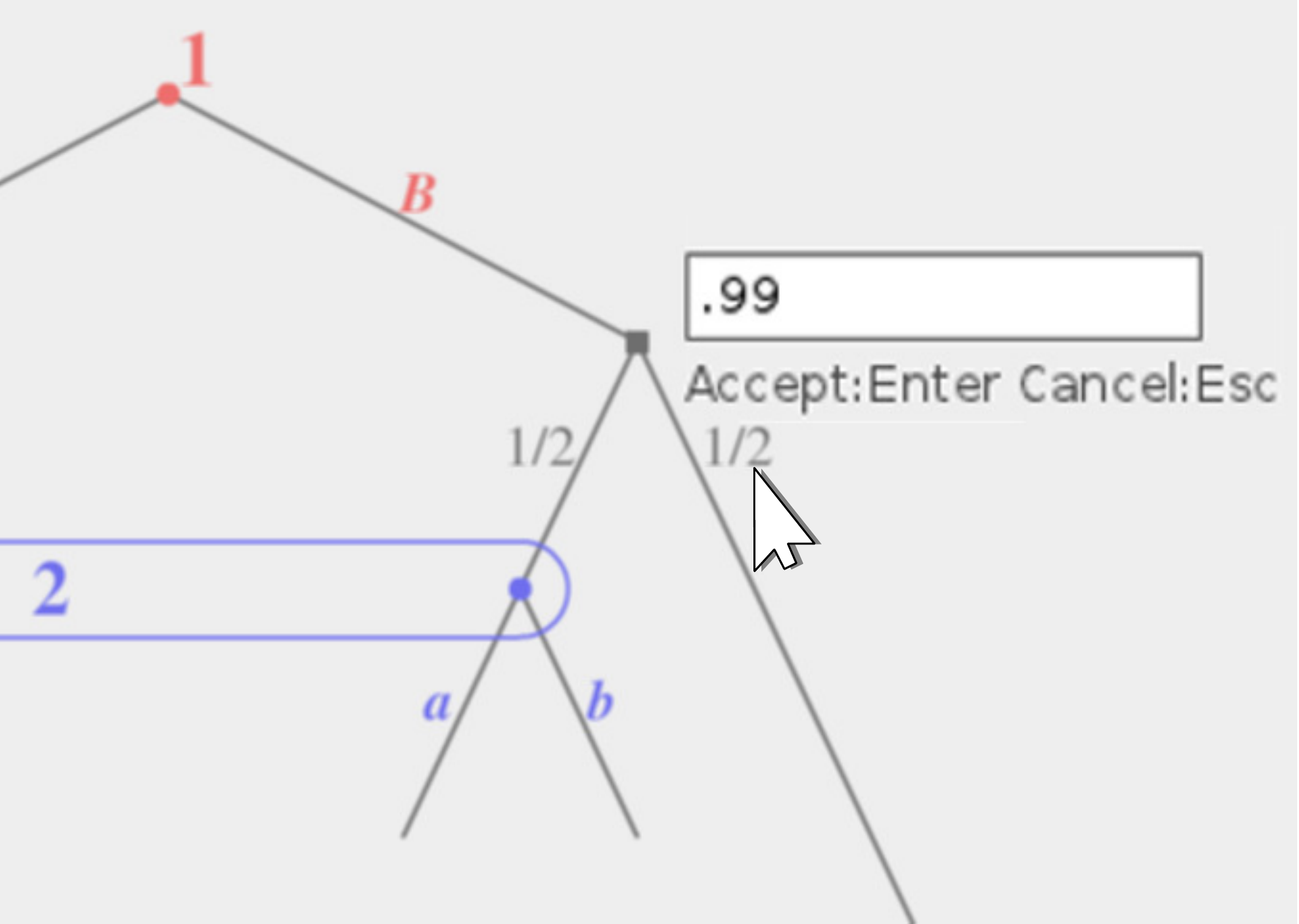}}
\hfill\strut
\includegraphics[height=14\emm]{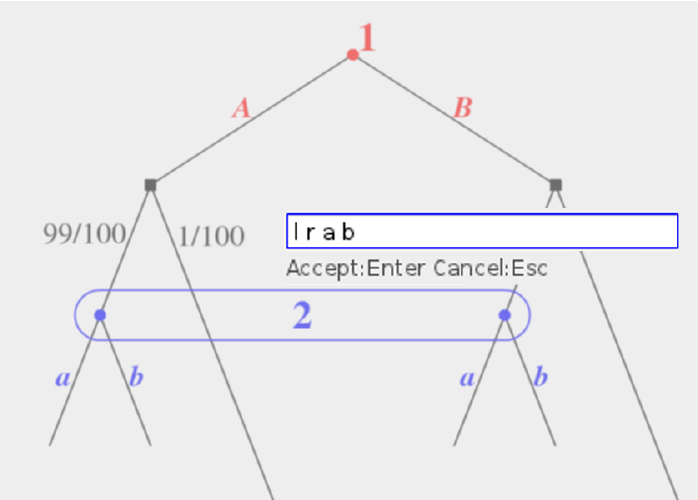} 
\caption{Changing move names, either by clicking on a single
move name or chance probability (left), or by changing all
moves of a player at once (right).}
\label{fmoves}
\end{figure}

Fig.~\ref{fpayoffs} shows the game tree after the
completed stages of assigning moves (and chance
probabilities) and payoffs, which work as follows.
When the ``moves'' stage is chosen in the headline bar,
all outgoing edges at an information set get unique
pre-assigned names from a list by traversing the tree in
``breadth-first'' manner.
Per default these are the upper-case letters in alphabetical
order for player~1 and the lower-case letters for player~2.
There are two ways to change these default move names:
First, by clicking on a move label, which allows to enter an
alternative name via the keyboard; this is not restricted to
a single letter and need not be unique.
On the left in Fig.~\ref{fmoves}, this is shown for the
chance probability of the right chance move which is changed
from $1/2$ to $0.99$ (which could also be entered as
$99/100$).
The remaining probability for the other chance move is
automatically set to $0.01$ so that the probabilities sum to
one.
A second, quick way to change all move names is the
mode button below the ``players'' stage in the headline bar
where all move names for a player are entered at once,
or altered from the displayed current list; this is shown in
the right picture of Fig.~\ref{fmoves} for player~2 whose
default moves $a,b,c,d$ are changed to $l,r,a,b$.  

\begin{figure}[hbt]
\strut \hfill
\includegraphics[height=14\emm]{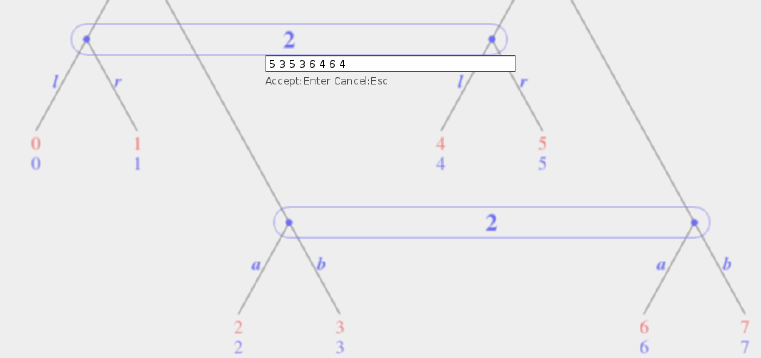} 
\hfill \strut 
\caption{Changing all payoffs of player~1 from the pre-set
values 0 1 2 3 4 5 6 7 (for easy identification) to their
intended values.} 
\label{feditpayoffs}
\end{figure}

At the final ``payoffs'' stage, the leaves of the tree get
payoffs to the two players, which are at first consecutively
given as $0,1,2,\ldots$ in order to have a unique payoff to
identify each leaf.
As shown in Fig.~\ref{feditpayoffs}, a player's payoffs
can then be changed at once by replacing them with the
intended payoffs for the game, where the numbering helps
to identify the leaves.
Payoffs can also be changed individually.
In addition, payoffs can be generated randomly.
A zero-sum option can be chosen which automatically sets the
other player's payoff to the negative of the payoff to the
current player.  

The game tree is stored as its logical structure.
Its graphical layout is generated automatically.
Its parameters can be changed, such as the orientation of
the game tree, which can grow vertically or horizontally in
either direction, the default being top-down.
Fig.~\ref{fsettings} shows a change of settings so that
the game tree grows from left to right.
Other parameters such as the colors used for the players,
line thickness, fonts, and other dimensions can also be
changed.  

Fig.~\ref{fsettings} also shows the general layout of the
graphical interface in the web browser.
The top left offers file manipulation functions such as
starting a new game, storing and loading the current game
(together with its settings), and exporting it to a picture
format (.png) and a scalable graphics format (.fig) that can
be further manipulated with the \verb.xfig. drawing program
and converted to .pdf or .eps files for inclusion in
documents.
On the top right various solution algorithms can be selected
and started, and at the very right zoom operations and
change of settings can be selected.

\begin{figure}[hbt]
\strut \hfill
\includegraphics[height=11\emm]{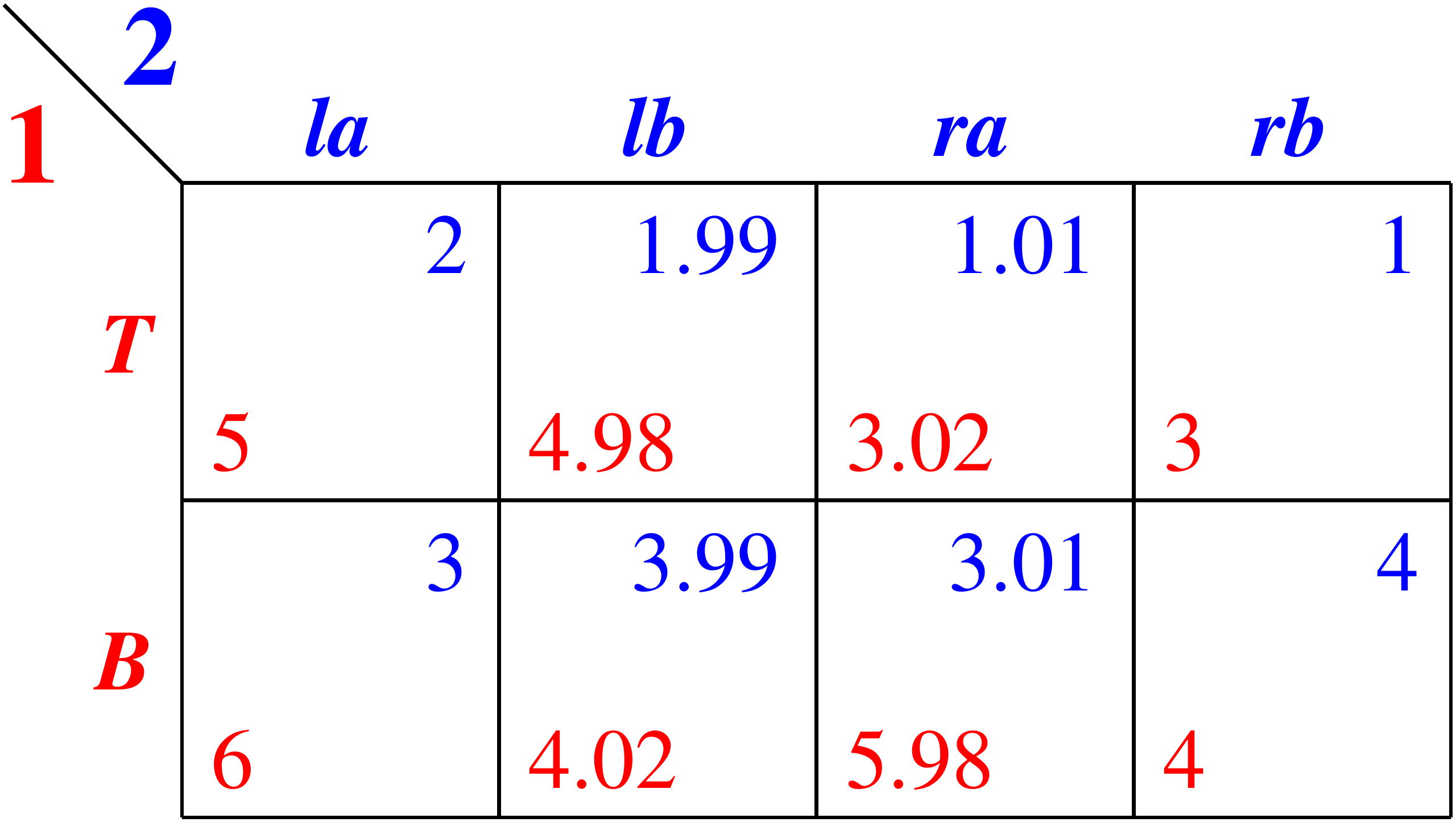}
\hfill\strut
\caption{Strategic form of the game in Fig.~\ref{fpayoffs}.}
\label{fimpsf}
\end{figure}

The game in Fig.~\ref{fpayoffs} has an interesting
equilibrium structure.
It is due to Bagwell (1995) and represents the commitment
game shown earlier in Fig.~\ref{f3}, but where the
commitment of player~1 is imperfectly observed due to some
transmission noise:
Namely, with a small probability (here $0.01$), the second
player observes the commitment incorrectly as the opposite
move, which is represented by the chance moves and the two
information sets of player~2.
The resulting strategic form is shown in
Fig.~\ref{fimpsf}.
Finding all equilibria with GTE gives the following output,
with three equilibria that are isolated, each in a separate
component.

{\footnotesize
\begin{verbatim}
Strategic form: 
2 x 4 Payoff player 1

  la     lb     ra rb
T  5 249/50 151/50  3
B  6 201/50 299/50  4

2 x 4 Payoff player 2

  la      lb      ra rb
T  2 199/100 101/100  1
B  3 399/100 301/100  4

EE = Extreme Equilibrium, EP = Expected Payoffs

Rational:
EE 1 P1: (1)  1/100 99/100 EP= 393/98 P2: (1)     0 25/49 0 24/49 EP= 397/100 
EE 2 P1: (2)      0      1 EP=      4 P2: (2)     0     0 0     1 EP=       4 
EE 3 P1: (3) 99/100  1/100 EP= 489/98 P2: (3) 24/49 25/49 0     0 EP= 201/100 

Decimal:
EE 1 P1: (1) 0.01 0.99 EP= 4.0102 P2: (1)      0 0.5102 0 0.4898 EP= 3.97 
EE 2 P1: (2)    0  1.0 EP=    4.0 P2: (2)      0      0 0    1.0 EP=  4.0 
EE 3 P1: (3) 0.99 0.01 EP= 4.9898 P2: (3) 0.4898 0.5102 0      0 EP= 2.01 

Connected component 1:
{1}  x  {1}

Connected component 2:
{2}  x  {2}

Connected component 3:
{3}  x  {3}
\end{verbatim}}

In particular, the original SPE $(T,lb)$ of the commitment
game with perfect observation and payoffs $5$ and $2$ has
disappeared.
Because the game does not have any nontrivial subgames
(which are subtrees where each player knows that they are in
the subgame), the concepts of subgame perfection and
backward induction no longer apply in the game with
imperfectly observable commitment in Fig.~\ref{fpayoffs}.
The reason why $(T,lb)$ is no longer an equilibrium is the
following.
Because player~1 commits to move $T$ with certainty,
player~2 should choose move~$l$ when seeing $T$, because
this gives her a higher payoff than~$r$.
However, when player~2 sees move $B$ of player~1, it must be
due to an error in the observation and player~2 would
therefore also choose $a$ rather than $b$; in short, $la$
and not $lb$ is a best response to $T$.
However, $(T,la)$ is not an equilibrium either because
against $lb$ player~1 would choose $B$; in a sense, player~1
would exploit the fact that player~2 interprets $B$ as an
erroneous signal.
This, however, seems to imply that player~1 has lost his
commitment power due to the noise in the observation.

A Bagwell (1995) pointed out, and as is shown in the above
list of equilibria, this loss of commitment power only
applies when the players are restricted to use pure
strategies.
There is in fact a mixed equilibrium, listed as the
component \ \verb.{3} x {3}.\,, which has payoffs $489/98$ and
$201/100$ to the two players that are close to the
``Stackelberg'' payoffs $5$ and $2$ when no noise is
present.
Here player~1 himself adds a small amount of noise to the
commitment and plays $T$ and~$B$ with probabilities $0.99$
and $0.01$.
In turn, player~2 mixes between $la$ and $lb$ with
probabilities $24/49$ and $25/49$.
That is, player~2 chooses $l$ with certainty and is
indifferent between $a$ and~$b$ because when she sees move
$B$, this signal may equally likely be received due to
the chance move or due to player~1's randomization.
The mixture between $a$ and~$b$ is such that player~1 in
turn is indifferent between $T$ and~$B$.
The game also has the pure strategy equilibrium $(B,rb)$ 
as before, listed as \ \verb.{2} x {2}. \ with payoffs $4$
and~$4$, and another mixed equilibrium \ \verb.{1} x {1}. \
with similar payoffs.

With GTE, the game in Fig.~\ref{fpayoffs} is created in
a few minutes, and the equilibria are computed instantly.
The game does not allow abstract parameters, such as
$\varepsilon$ for the error probability as in the analysis
by Bagwell (1995), which is here chosen as~$0.01$.
However, as a quick way to test a typical case of this
game-theoretic model, GTE is a valuable tool.

\section{Examples of analyzing games in strategic form}
\label{s-strategic}

In this section we give examples of strategic-form games
analyzed with GTE.
These also demonstrate the use of GTE as a research tool for
game theory, for questions on the possible number of
equilibria, or the description of equilibrium components in
the context of strategic stability.

\begin{figure}[hbt]
\strut \hfill
\includegraphics[width=35\emm]{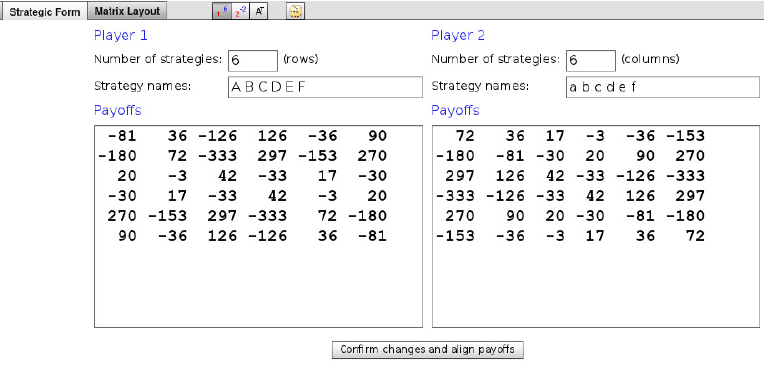}
\hfill\strut
\caption{GTE input of a $6\times6$ game in strategic form.
}
\label{fstratinput}
\end{figure}

Fig.~\ref{fstratinput} shows the input of a game in
strategic form, currently implemented for two players only.
For each player one needs to specify the number of
strategies, which are the rows for player~1 and columns for
player~2, and a payoff matrix.
Names for the strategies are generated automatically as
upper case letters for player~1 and lower case letters for
player~2, and can be changed.
If the strategic form has been generated from the extensive
form, then the strategies are shown as tuples of moves, one
for each information set.

One can choose a {\em zero-sum} input mode where the payoffs
to player~2 are automatically the negative of the payoffs to
player~1.
Similarly, one can input a {\em symmetric} game where the
square payoff matrices to player 1 and~2 are a matrix $A$
and its transpose~$A\T$; the game is symmetric because it
does not change when the players are exchanged.
In both cases, only the payoff matrix of player~1 is
entered.

Payoffs can be entered as integers, fractions, or with a
decimal point where the display can be switched between
fractions and decimals; internally they are all stored as
fractions.
The ``Align and Update'' button aligns the entries in
columns and updates the second player's payoffs for zero-sum
and symmetric games.

The input of bimatrix games and the computation of their
equilibria has the functionality of the popular webpage of
Savani (2005) which has been used tens of thousands of
times.
The extra feature in GTE is the graphical output as in
Fig.~\ref{fimpsf}, for example, which is accessed by the
``Matrix Layout'' tab shown at the top of
Fig.~\ref{fstratinput}.

The game in Fig.~\ref{fstratinput} has 75 equilibria,
listed as follows; the output displays each equilibrium in a single
line, which we have here broken into two lines, one per
player, to fit the page.

{\footnotesize
\begin{verbatim}
    EE  1 P1:  (1)   1/30    1/6    3/10    3/10    1/6   1/30 EP=     3/2
          P2:  (1)    1/6   1/30    3/10    3/10   1/30    1/6 EP=     3/2 
    EE  2 P1:  (2)      0      0    1/33    5/33   4/11   5/11 EP=   24/11
          P2:  (2)      0      0    5/33    1/33   5/11   4/11 EP=   24/11 
    EE  3 P1:  (3)  1/128      0       0    7/64 47/128  33/64 EP=    12/7
          P2:  (3)      0      0    5/21  13/189 59/189   8/21 EP= 297/128 
    ...
    EE 74 P1: (74)      0      0   13/32       0      0  19/32 EP= 450/103
          P2: (74) 33/103 70/103       0       0      0      0 EP=  477/16 
    EE 75 P1: (75)      0      0       0       0      1      0 EP=     270
          P2: (75)      1      0       0       0      0      0 EP=     270 

    Connected component 1:
    {1}  x  {1}
    ...
    Connected component 75:
    {75}  x  {75} 
\end{verbatim}}

This game is the smallest known example that refutes a
conjecture by Quint and Shubik (1995) that an $n\times n$
game has at most $2^n-1$ equilibrium components, here for
$n=6$.
It has been constructed by von Stengel (1999) using methods
from polytope theory, with the specific small integers in 
Fig.~\ref{fstratinput} described by Savani and von Stengel
(2004, p.~25).
Studying games with large numbers of equilibria is obviously
greatly aided by computational tools; for another example
see von Stengel (2012).

\begin{figure}[hbt]
\begin{minipage}{.295\hsize}
\includegraphics[width=10.8\emm]{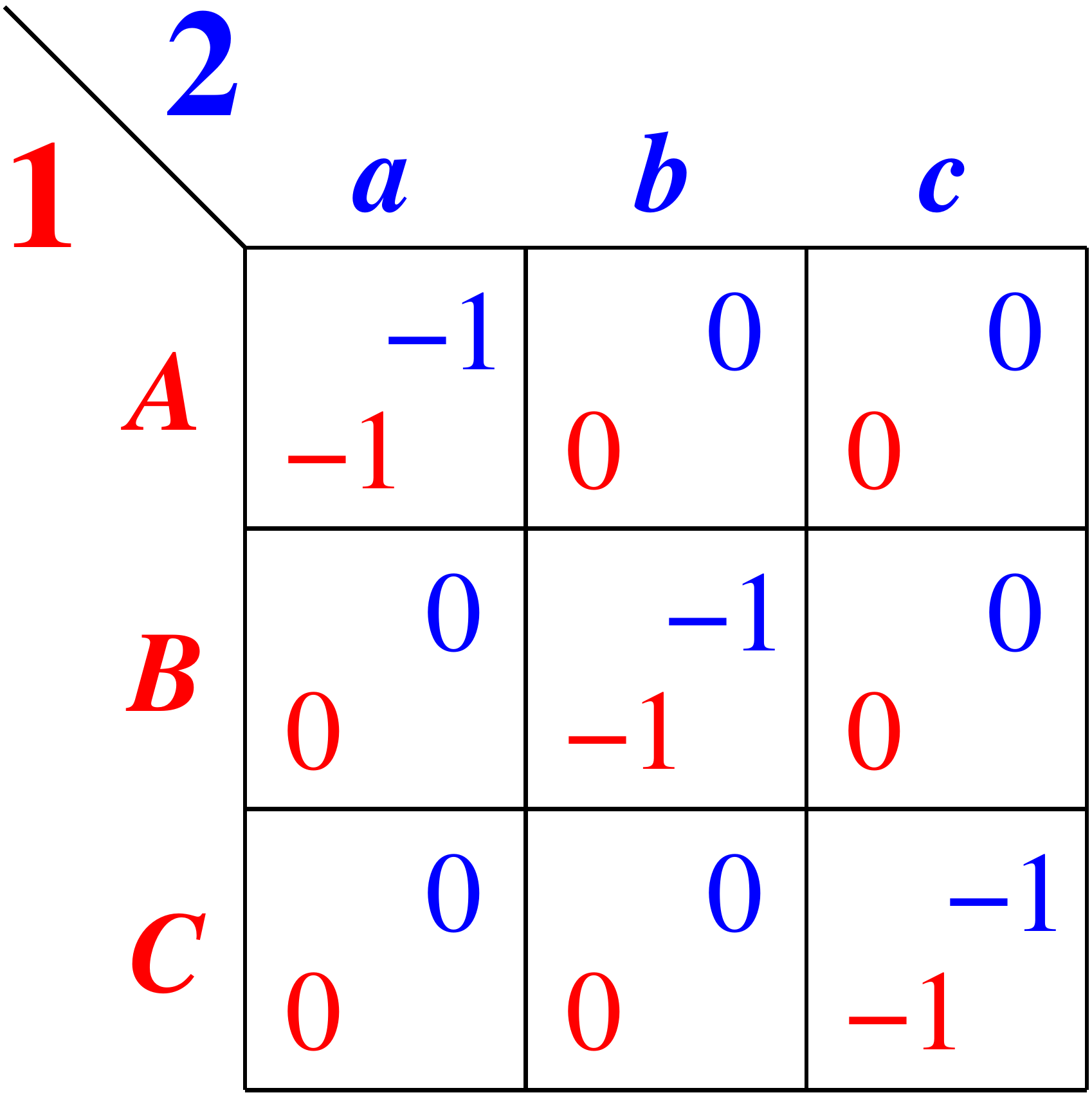}
\vskip4ex
\includegraphics[width=10.8\emm]{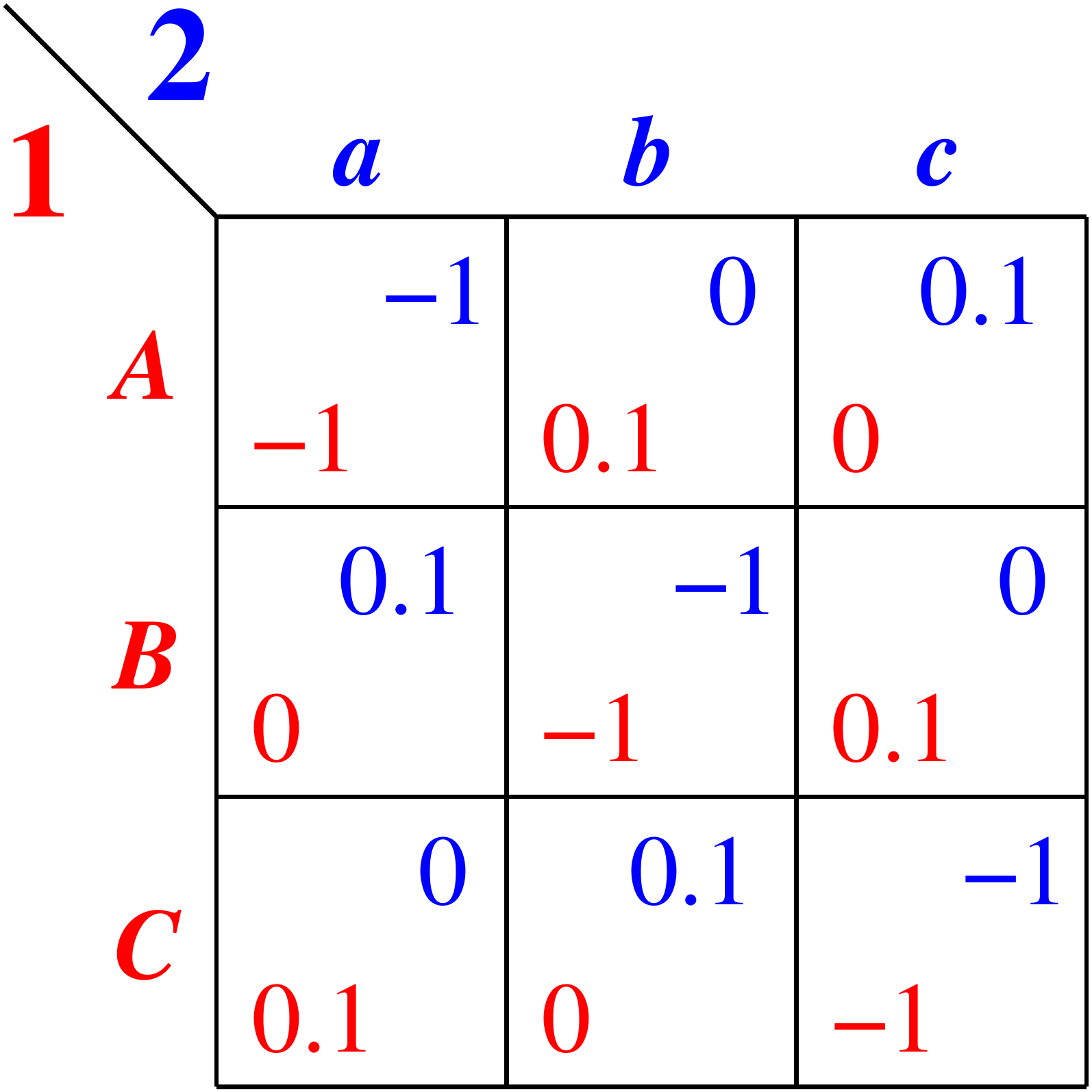} 
\end{minipage}%
\begin{minipage}{.682\hsize}
\scriptsize
\begin{verbatim}
EE 1 P1: (1) 1/3 1/3 1/3 EP= -1/3 P2: (1) 1/3 1/3 1/3 EP= -1/3 
EE 2 P1: (2)   1   0   0 EP=    0 P2: (2)   0   1   0 EP=    0 
EE 3 P1: (2)   1   0   0 EP=    0 P2: (3)   0   0   1 EP=    0 
EE 4 P1: (3)   0   1   0 EP=    0 P2: (3)   0   0   1 EP=    0 
EE 5 P1: (3)   0   1   0 EP=    0 P2: (4)   1   0   0 EP=    0 
EE 6 P1: (4)   0   0   1 EP=    0 P2: (4)   1   0   0 EP=    0 
EE 7 P1: (4)   0   0   1 EP=    0 P2: (2)   0   1   0 EP=    0 

Connected component 1:
{1}  x  {1}

Connected component 2:
{2, 4}  x  {2}
{3, 4}  x  {4}
{4}  x  {2, 4}
{2, 3}  x  {3}
{3}  x  {3, 4}
{2}  x  {2, 3} 

----------------------------------------------------------------

EE 1 P1: (1) 1/3 1/3 1/3 EP= -3/10 P2: (1) 1/3 1/3 1/3 EP= -3/10

Connected component 1:
{1}  x  {1} 
\end{verbatim}
\end{minipage} 
\caption{Top left: $3\times3$ game which has two equilibrium
components, shown on the right.
Bottom left: The same game but with {\em perturbed} payoffs
so that only component~1 remains (see bottom right).
}
\label{fring}
\end{figure}

Fig.~\ref{fring} shows at the top left a symmetric
``anti-coordination'' game where the only nonzero payoffs
are $-1$ to both players on the diagonal.
Because no payoff is positive, any cell with payoff zero to
both players is an equilibrium, and these equilibria are
connected by line segments to form a ``ring'' that defines
the topologically connected component~2 as in the output
shown on the right in Fig.~\ref{fring}.
The game also has an isolated completely mixed equilibrium
shown as component~1.
Interestingly, only component~1 is {\em strategically
stable} in the sense that there is always an equilibrium
nearby when the payoffs are slightly perturbed.
In the game shown at the bottom left, the payoffs are
perturbed so that the only equilibrium that remains
is the completely mixed equilibrium in component~1;
the perturbations from zero to $0.1$ can be changed to
independent arbitrarily small positive reals with the same
effect.

\begin{figure}[hbt]
\begin{minipage}{.40\hsize}
\includegraphics[width=12\emm]{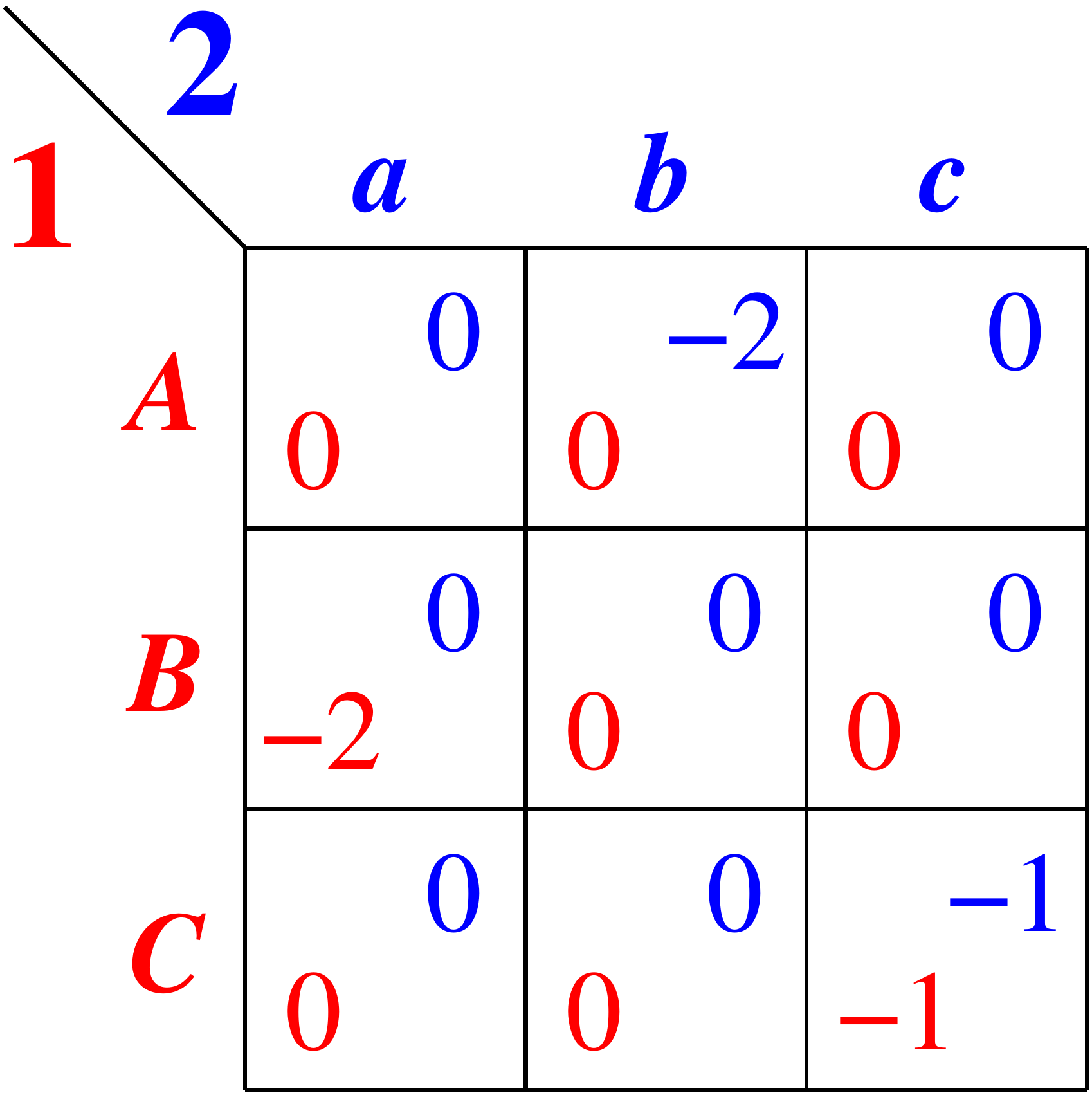}

\strut
\end{minipage}
\begin{minipage}{.50\hsize}
\scriptsize
\begin{verbatim}
EE 1 P1: (1) 0 1 0 EP= 0 P2: (1) 0 1 0 EP= 0 
EE 2 P1: (1) 0 1 0 EP= 0 P2: (2) 0 0 1 EP= 0 
EE 3 P1: (2) 1 0 0 EP= 0 P2: (3) 1 0 0 EP= 0 
EE 4 P1: (2) 1 0 0 EP= 0 P2: (2) 0 0 1 EP= 0 
EE 5 P1: (3) 0 0 1 EP= 0 P2: (3) 1 0 0 EP= 0 
EE 6 P1: (3) 0 0 1 EP= 0 P2: (1) 0 1 0 EP= 0 

Connected component 1:
{1, 3}  x  {1}
{2, 3}  x  {3}
{3}  x  {1, 3}
{1, 2}  x  {2}
{2}  x  {2, 3}
{1}  x  {1, 2} 
\end{verbatim}
\end{minipage} 
\caption{$3\times3$ game which has only a ``ring'' of
equilibria as its sole stable component, shown on the right.
}
\label{fKM}
\end{figure}

The concept of strategic stability is due to Kohlberg and
Mertens (1986).
Fig.~\ref{fKM} shows a game that is also symmetric like
that at the top left of Fig.~\ref{fring} (the symmetry is
easy to see due to the staggered payoffs in the lower left
and upper right of each cell).
It is strategically equivalent (by subtracting constants
from columns of the row player's payoffs and from rows of
the column player's payoffs) to the game of Kohlberg and
Mertens (1986, p.~1034) and has a ``ring'' consisting of
line segments as its only equilibrium component.
As Kohlberg and Mertens have shown, any point on that ring
can be chosen so that a suitably perturbed game has its only
equilibrium near that point, which can be verified by
experimenting with numeric perturbations of the game.
Correspondingly, the entire ring defines the minimal stable
component of the game.

\begin{figure}[hbt]
\begin{minipage}{.325\hsize}
\includegraphics[width=12\emm]{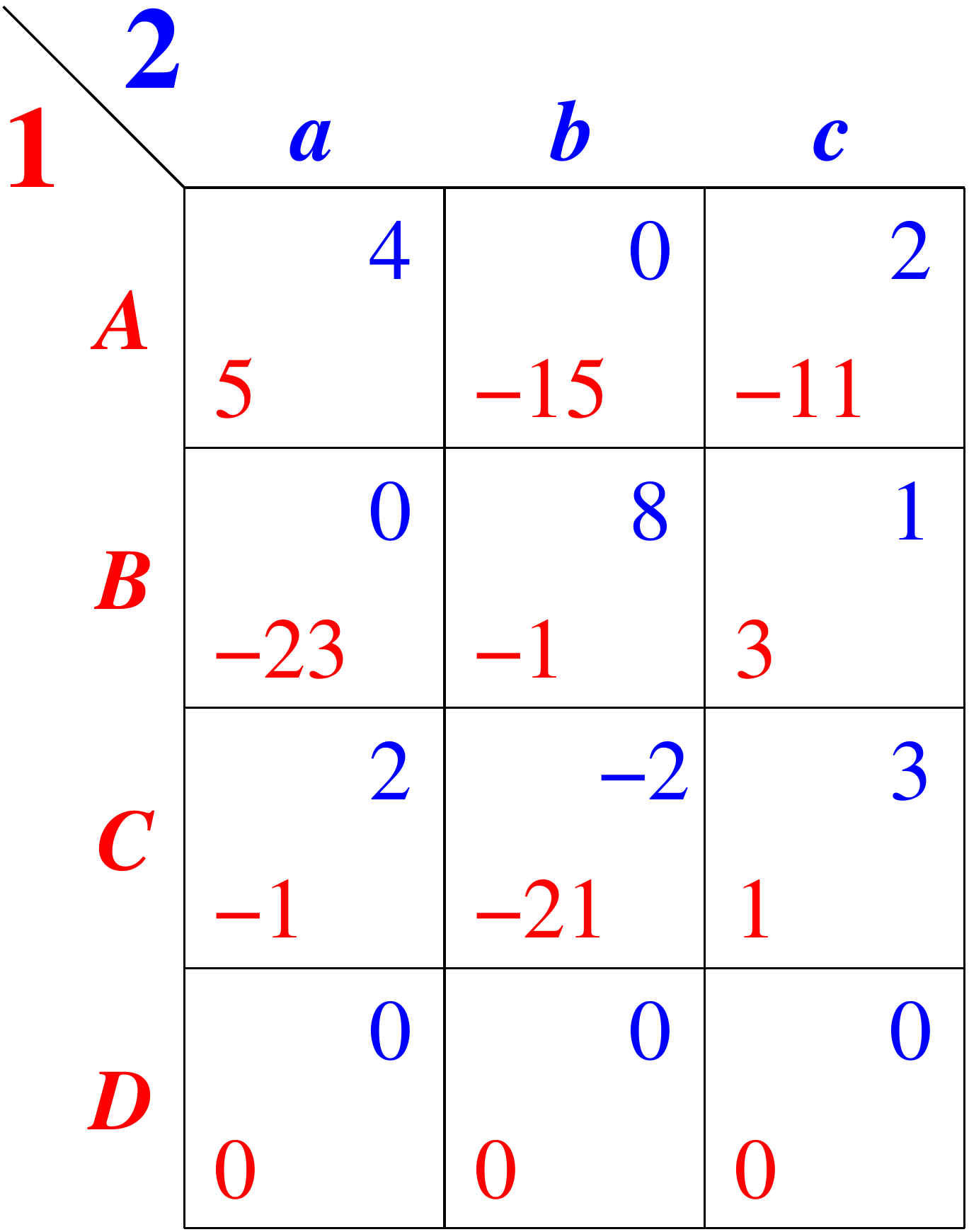}
\vskip2ex
\strut
\end{minipage} 
\begin{minipage}{.67\hsize}
\scriptsize
\begin{verbatim}
EE 1 P1: (1) 0 0 0 1 EP= 0 P2: (1) 31/282 5/141 241/282 EP= 0 
EE 2 P1: (1) 0 0 0 1 EP= 0 P2: (2)      0   3/4     1/4 EP= 0 
EE 3 P1: (1) 0 0 0 1 EP= 0 P2: (3)      0     1       0 EP= 0 
EE 4 P1: (1) 0 0 0 1 EP= 0 P2: (4)    3/4   1/4       0 EP= 0 
EE 5 P1: (1) 0 0 0 1 EP= 0 P2: (5)    1/2     0     1/2 EP= 0 
EE 6 P1: (1) 0 0 0 1 EP= 0 P2: (6)  11/16     0    5/16 EP= 0 
EE 7 P1: (2) 1 0 0 0 EP= 5 P2: (7)      1     0       0 EP= 4 

Connected component 1:
{1}  x  {1, 2, 3, 4, 5, 6}

Connected component 2:
{2}  x  {7} 
\end{verbatim}
\vskip-16ex
\strut\hfill
\includegraphics[width=.5\hsize]{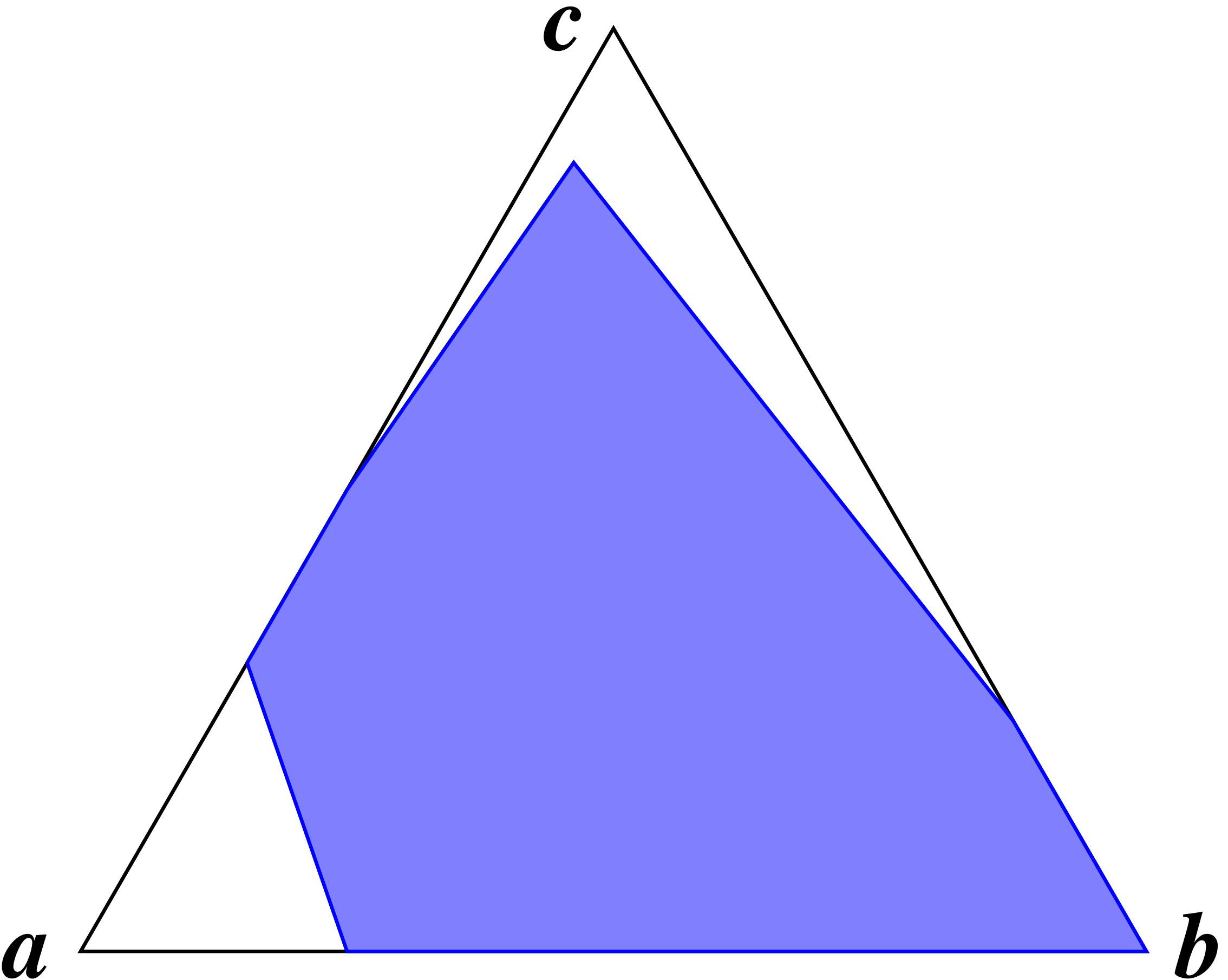} 
\end{minipage} 
\caption{Left: $4\times3$ game with two equilibrium
components, shown on the right.
In component~1, player~1 plays 
$D$
and player~2 plays so that player~1 gets at most payoff~$0$,
shown by the hexagon at the bottom right (see also
Hauk and Hurkens, 2002, Fig.~4).
Component~2 is the pure-strategy equilibrium $(A,a)$.
}
\label{fhaukhurkens}
\end{figure}

Fig.~\ref{fhaukhurkens} shows another game related to
strategic stability, due to Hauk and Hurkens (2002, p.~74).
This game corresponds to an extensive game where player~1
can first choose between an ``outside option'' $D$ with
constant payoff zero to both players or else play in a
$3\times 3$ simultaneous game.
The game has two equilibrium components, the pure
equilibrium $(A,a)$ and the outside option component where
player~2 randomizes in such a way that player~1 gets at most
payoff zero and therefore will not deviate from~$D$.

Interestingly, the larger component in this game has the
property that it has {\em index} zero but is nevertheless
strategically stable, that is, any perturbed game has
equilibria near that component, as shown by Hauk and Hurkens
(2002).
The larger component in the game in Fig.~\ref{fring} has
also index zero and is not stable, which is what one would
normally expect.
The index is a certain integer defined for an equilibrium
component, with the important property that the sum over all
equilibrium indices is one, and that for generic
perturbations each equilibrium has index $1$ or~$-1$, which
is always $1$ for a pure-strategy equilibrium
(see Shapley, 1974).
A symbolic computation of the index of an equilibrium
component for a bimatrix game is described by Balthasar
(2009, Chapter~2). 
Its implementation is planned as a future additional feature
of GTE.

All currently implemented algorithms of GTE apply only to
two-player games.
A simple next stage is to allow three or more players at
least for the purpose of drawing games and generating their
strategic form.
Algorithms for finding all equilibria of a game with any
number of players using polynomial algebra are described by
Datta (2010); some of these require computer algebra
packages, so far not part of GTE.

\section{Equilibrium computation for games in strategic form}
\label{s-algostrat}

In this section, we describe the algorithm used in GTE that
finds all equilibria of a two-player game in strategic form.
This algorithm searches the vertices of certain polyhedra
defined by the payoff matrices, which is practical for up to
about 20 strategies per player.
For larger games, one can normally still find in reasonable
time one equilibrium, or possibly several equilibria with
varying starting points, using the path-following algorithms
by Lemke and Howson (1964) and Lemke (1965) mentioned at the
end of this section.
We assume some familiarity with {\em pivoting} as used in
the simplex algorithm for linear programming, which in the
present context is explained further in Avis et al.\ (2010).

The {\em best response condition} describes, in a general
finite game in strategic form, when a profile of mixed
strategies is a Nash equilibrium.
If player~1, say, has $m$ pure strategies, then his set $X$
of mixed strategies is the set of probability
vectors $x=(x_1,\ldots,x_m)$ so that $x_i\ge0$ and
$\sum_{i=1}^mx_i=1$.
Geometrically, $X$ is a {\em simplex} which is the convex
hull of the $m$ unit vectors in $\reals^m$.
If $x$ is part of a Nash equilibrium, then $x$ has to be a
best response against the strategies of the other players.
The best response condition states that this is the case if
and only if every {\em pure strategy} $i$ so that $x_i>0$ is
such a best response (the condition is easy to see and was
used by Nash, 1951).
This is a finite condition, which requires to compute
player~1's expected payoff for each pure strategy~$i$, and
to check if the strategies $i$ in the {\em support} 
$\{i\mid x_i>0\}$ of~$x$ give indeed maximal payoff.
These maximal payoffs have to be {\em equal}, and the
resulting equations typically determine the probabilities of
the mixed strategies of the other players.

\begin{figure}[hbt]
\begin{minipage}{.515\hsize}
\leavevmode
\includegraphics[height=11\emm]{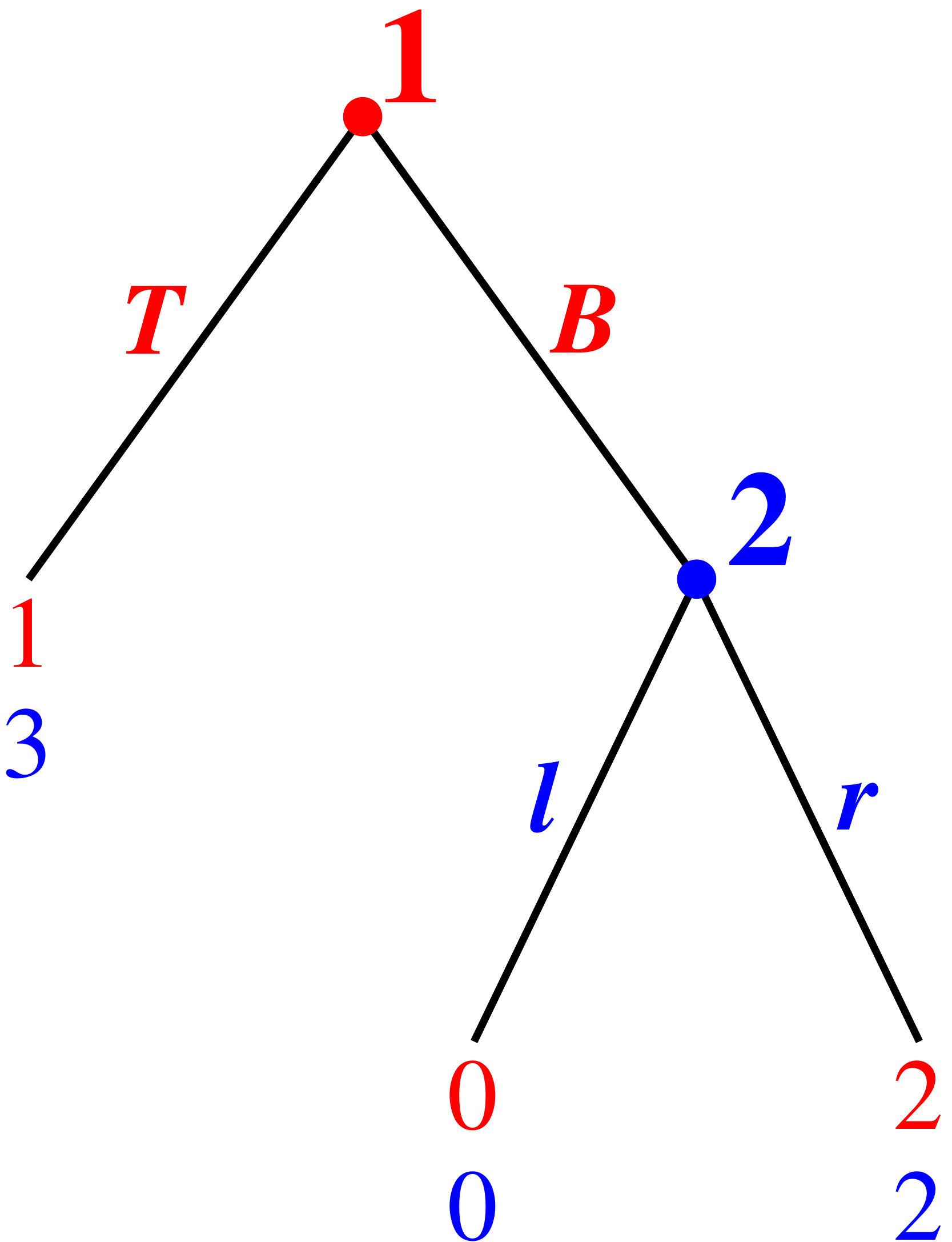} 
\hfill
\raise2\emm\hbox{\includegraphics[height=8\emm]{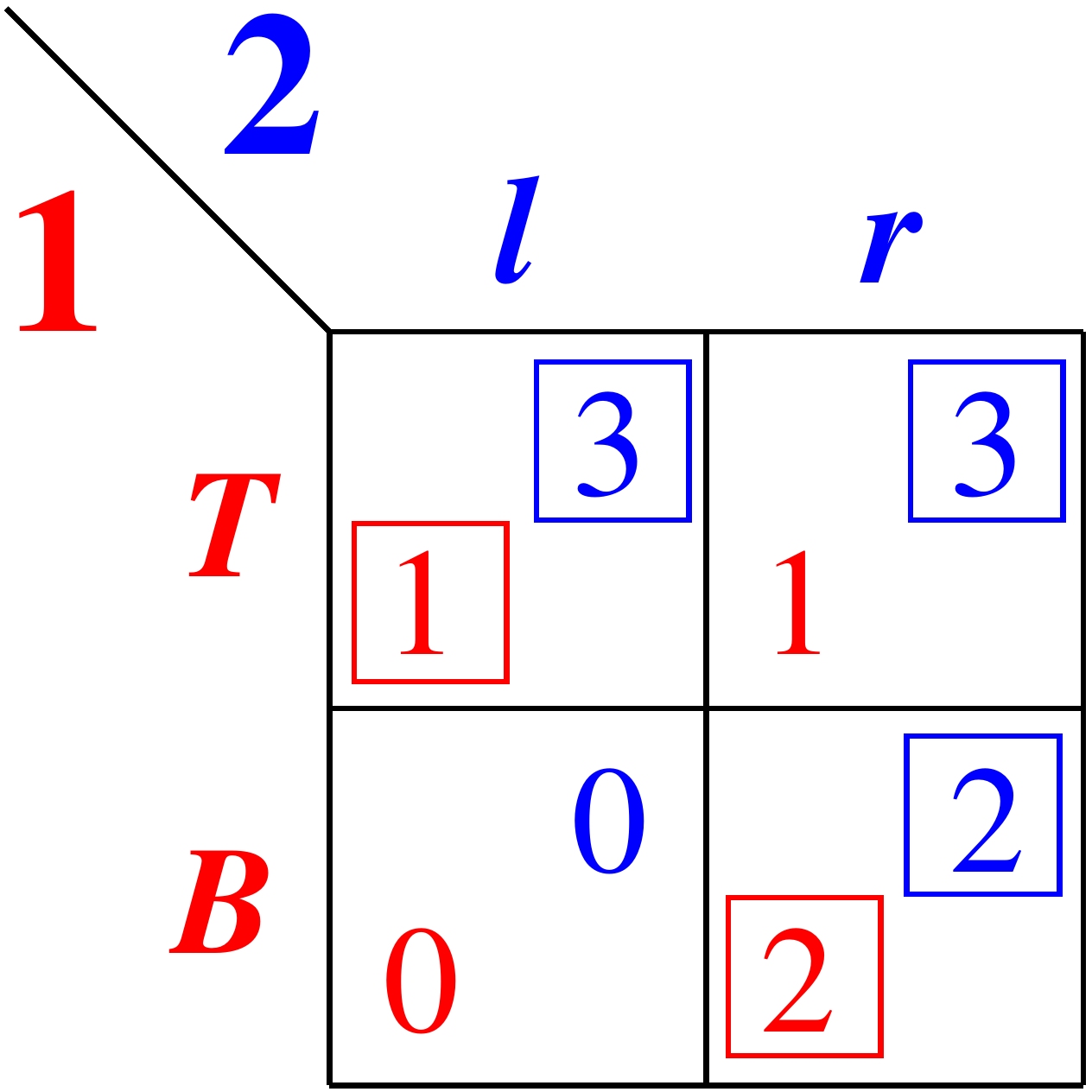}}
\hfill
\hfill
\strut
\end{minipage}%
\begin{minipage}{.485\hsize}
\scriptsize
\begin{verbatim}
EE 1 P1: (1) 1 0 EP= 1 P2: (1)   1   0 EP= 3 
EE 2 P1: (1) 1 0 EP= 1 P2: (2) 1/2 1/2 EP= 3 
EE 3 P1: (2) 0 1 EP= 2 P2: (3)   0   1 EP= 2 

Connected component 1:
{1}  x  {1, 2}

Connected component 2:
{2}  x  {3} 

\end{verbatim}
\end{minipage} 
\caption{A ``threat game'' in extensive form, its $2\times 2$
strategic form, and its equilibrium components.
}
\label{fthreat}
\end{figure}

For two players, these constraints define ``best response
polyhedra'' that simplify the search for equilibria.
We illustrate this geometric apprach with an example; for
a general exposition and detailed references see von Stengel
(2002; 2007).  
Fig.~\ref{fthreat} shows a simple game tree which has 
the SPE $(B,r)$, and another equilibrium component where
player~2 ``threatens'' to choose $l$ with probability at
least $1/2$ and player~1 chooses~$T$.

\begin{figure}[hbt]
\strut \hfill
\includegraphics[width=.75\hsize]{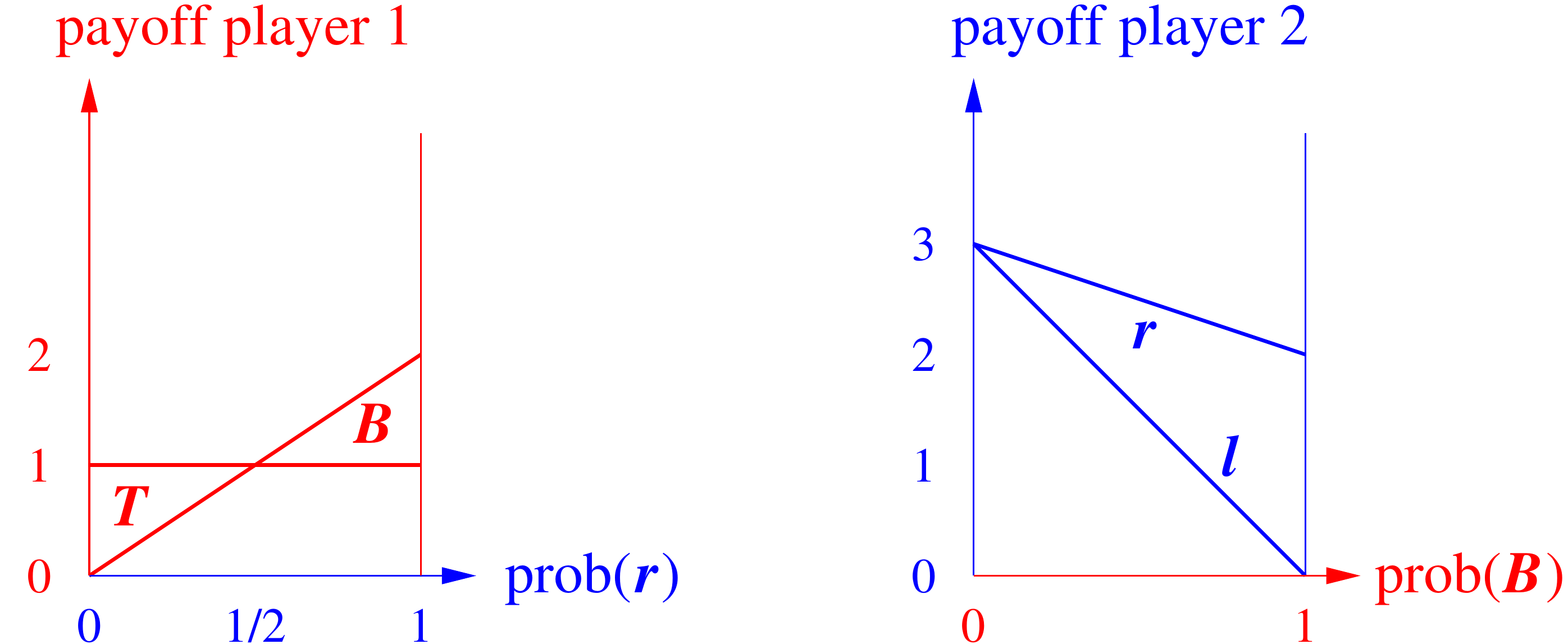} 
\hfill\strut
\caption{Expected payoffs to the two players in the threat
game in Fig.~\ref{fthreat} as a function of the mixed
strategy of the other player.
}
\label{fupper}
\end{figure}

Fig.~\ref{fupper} shows on the left the mixed strategy $y$
of player~2, which is defined by the probability that she
chooses move~$r$, say, together with the expected payoff to
player~1 for his two pure strategies $T$ and~$B$.
These expected payoffs are linear functions of $y$
(because the game has only two players; for more than two
players, the expected payoff for a pure strategy of player~1
is a product of mixed strategy probabilities of the other
players, which is no longer linear).
The right picture shows the expected payoffs for the pure
strategies $l$ and $r$ of player~2 as a function of the
mixed strategy of player~1.

These pictures show when a pure strategy is a {\em best
response} against a mixed strategy of the other player:
$T$ is a best response of player~1 when
prob$(r)\le 1/2$, and $B$ is a best response when
prob$(r)\ge1/2$.
Strategy $r$ of player~2 is always a best response,
and $l$ is a best response when prob$(B)=0$.

\begin{figure}[hbt]
\strut \hfill
\includegraphics[width=.75\hsize]{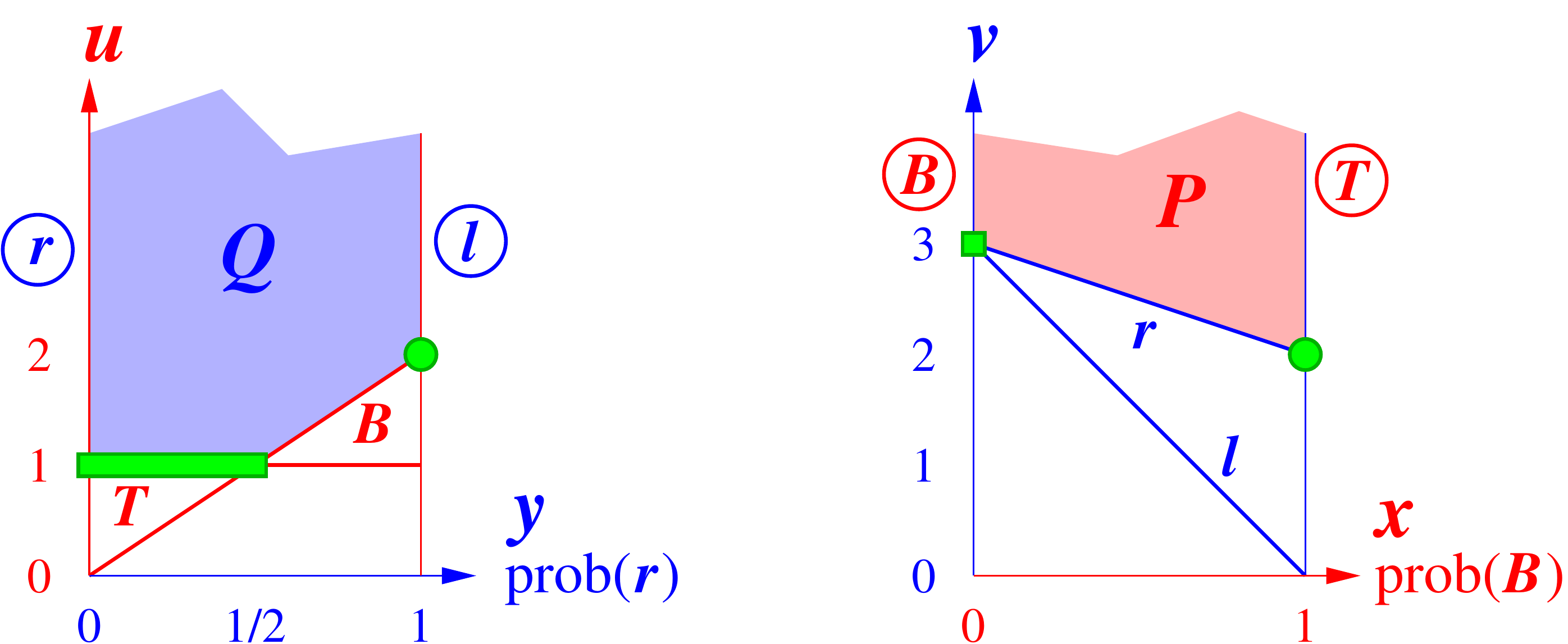} 
\hfill\strut
\caption{The polyhedra $Q$ and $P$ in (\ref{PQ}) that define
the ``upper envelope'' of the payoffs in
Fig.~\ref{fupper}, with additional circled facet labels
for unplayed strategies, and equilibrium strategy pairs.
}
\label{fupperlabel}
\end{figure}

In Fig.~\ref{fupperlabel}, the same pictures are shown more
abstractly where $u$ and $v$ are the expected payoffs to
player 1 and~2, respectively, which are required to be
at least as large as the expected payoff for every pure
strategy.
In a general $m\times n$ bimatrix game, let $a_{ij}$ and
$b_{ij}$ be the payoffs to player 1 and~2, respectively,
for each pure strategy pair~$i,j$.
Analogous to $X$, let $Y$ be the mixed strategy simplex of
player~2 given by $Y=\{y\in\reals^n\mid
y_j\ge0,~\sum_{j=1}^n y_j=1\}$.
Then $P$ and $Q$ are the {\em best response polyhedra}
\begin{equation}
\label{PQ}
\begin{array}{rcl}
P & = & \{(x,v)\in X\times\reals\mid
\sum_{i=1}^m b_{ij}x_i\le v,~1\le j\le n\},
\\
Q & = & \{(y,u)\in Y\times\reals\mid 
\sum_{j=1}^n a_{ij}y_j\le u,~1\le i\le m\}.
\end{array}
\end{equation}
In Fig.~\ref{fupperlabel}, $Q$ is shown on the left and
refers to the set of mixed strategies $y$ of player~2
together with the best responses of player~1 and
corresponding payoff~$u$ to player~1.
Consider one of the inequalities
$\sum_{j=1}^n a_{ij}y_j\le u$ in the definition of~$Q$ for
some pure strategy~$i$ of player~1, which in the example is
either $T$ or~$B$.
When this inequality is {\em tight}, that is, holds as
equality, then $i$ is clearly a best response to~$y$.
This tight inequality defines a {\em face}  (intersection
with a valid hyperplane) of the polyhedron, and is often a
facet (a face of maximum dimension), as here for both
strategies $T$ and~$B$;
however, for the polyhedron $P$ on the
right in Fig.~\ref{fupperlabel} the tight inequality
$\sum_{i=1}^m b_{ij}x_i=v$ when $j$ is the pure strategy~$l$
of player~2 does not define a facet because it only holds for
the single point where prob$(B)=0$ and $v=3$, that is, for
$(x,v)=((1,0),3)\in P$.

We {\em label} each point $(y,u)$ of $Q$ with the
strategy~$i$ of player~1 whenever $i$ is a best response
to~$y$, that is, when $\sum_{j=1}^n a_{ij}y_j=u$.
In addition, $(y,u)$ is labeled with strategy $j$ of
player~2 if $y_j=0$.
The labels for these {\em unplayed} pure strategies are
shown with circles around them in Fig.~\ref{fupperlabel}.
So any point $((1,0),u)$ where prob$(r)=0$ (the left edge of
$Q$) is labeled with $r$, and any point $((0,1),u)$ where
prob$(r)=1$ and hence prob$(l)=0$ (the right edge of~$Q$) is
labeled with~$l$.
Similarly, in the right picture the left and right edges of
$P$ have label $B$ and $T$, respectively.
The bottom edge of $P$ is labeled with the pure
strategy~$r$, and the left vertex $(x,v)=((1,0),3)$ has
label~$l$ in addition to the labels $B$ and~$r$.

With this labeling, an equilibrium $(x,y)$ of the game with
payoffs $u$ and $v$ to player 1 and~2 is given by a pair
$((x,v),(y,u))$ in $P\times Q$ that is {\em completely
labeled}, that is, each pure strategy, here $T,B,l,r$, of
either player appears as a label of $(x,v)$ or $(y,u)$.
Otherwise, a {\em missing label} represents a pure strategy
that has positive probability but is not a best response,
which does not hold in an equilibrium because it contradicts
the best response condition.

In Fig.~\ref{fupperlabel}, one equilibrium $(x,y)$,
indicated by the small disks to mark the two points $(x,v)$
and $(y,u)$, is given by $x=y=(0,1)$ with $u=v=2$, which is
the pure strategy pair $(B,r)$.
Here the point $(y,u)$ of $Q$ has labels $B$ and~$l$ and the
point $(x,v)$ of~$P$ has labels $r$ and~$T$, so these two
points together have all labels $T,B,l,r$.
A second equilibrium component, indicated by the rectangles,
is given by any point $(y,u)$ on the entire edge of $Q$ that
has label $T$ and $(x,v)=((1,0),3)$ in~$P$ which has the
three labels $r,l,B$.
These are also the two equilibrium components in the GTE
output in Fig.~\ref{fthreat}.
The edge of $Q$ with label~$T$ is the convex hull of the two
vertices $((1,0),1)$ with labels $T$ and~$r$, and
$((1/2,1/2),1)$ with labels $T$ and $B$.
Each of these vertices of~$Q$,
together with the vertex $((1,0),3)$ of $P$,
forms a completely labeled {\em vertex pair}.
For these two vertex pairs of $P\times Q$, one 
label ($r$ or $B$) appears twice and represents a
best-response pure strategy that is played with probability
zero, which is allowed in equilibrium.

The {\em extreme equilibria} computed by GTE are in fact all
vertex pairs of $P\times Q$ that are completely labeled,
which are then further processed so as to detect the
equilibrium components.
We give an outline of how these computations work in GTE,
described in detail and compared with other approaches by
Avis et al.\ (2010).

{\em Vertex enumeration}, that is, listing all vertices of a
polyhedron defined by linear inequalities, is a well-studied
problem where we use the {\em lexicographic reverse search}
method {\em lrs} by Avis (2000; 2006).
This method reverses the steps of the simplex algorithm for
linear programming for a certain deterministic pivoting rule.
It starts with a vertex $v_0$ of the polyhedron and a linear
objective function that is maximized at that vertex.
The simplex algorithm for maximizing this linear function
computes from every vertex a path of {\em pivoting steps}
to~$v_0$.
With a deterministic pivoting rule, that path is unique.
In lrs, the pivoting rule chooses as entering variable the
variable with the least index (i.e., smallest subscript)
that improves the objective function, and the leaving
variable via a lexicographic rule.
The unique paths of simplex steps from the vertices to
$v_0$ define a tree with root~$v_0$.
The lrs algorithm explores this tree by traversing the tree
edges in the reverse direction using a depth-first search.

In lrs, as in the simplex algorithm, the vertices of
$P\times Q$ are represented by {\em basic feasible
solutions} to the inequalities in (\ref{PQ}) when they are
represented in equality form via slack variables.
Here, these are vectors $s$ in $\reals^n$ and $r$ in
$\reals^m$ so that the constraints that define $P$ and $Q$
are written as
\begin{equation}
\label{rs}
\sum_{i=1}^m b_{ij}x_i+s_j= v, \qquad r_i+\sum_{j=1}^n
a_{ij}y_j= u, \qquad x_i,s_j,r_i,y_j\ge0
\end{equation}
for $1\le i\le m$, $1\le j\le n$, and $\sum_{i=1}^n x_i=1$
and $\sum_{j=1}^n y_j=1$ so that $x\in X$ and $y\in Y$.  A
feasible solution $x,s,v,r,y,u$ to (\ref{rs}) defines a
point in $P\times Q$ and vice versa.
A tight inequality corresponds to a slack variable that is
zero.
In fact, the labels for a strategy pair $x,y$, where $u$ and
$v$ are chosen minimally so that (\ref{rs}) holds, which
determines $r$ and $s$, are the pure strategies $i$ so that
$x_i=0$ or $r_i=0$ for player~1 and $y_j=0$ or $s_j=0$ for
player~2.
The equilibrium condition of being completely labeled is
equivalent to the {\em complementarity} condition
\begin{equation}
\label{compl}
x_i\,r_i=0 \quad(1\le i\le m),
\qquad y_j\,s_j=0 \quad(1\le j\le n).
\end{equation}
In a basic feasible solution to (\ref{rs}), the labels
correspond to the indices $i$ or $j$ of the 
nonbasic variables, and to basic variables that happen to
have value zero.

Basic solutions where basic variables have value zero are
called {\em degenerate}, and games where this happens are
also called degenerate.
It is easy to see (see von Stengel, 2002) that this
corresponds to a mixed strategy of a player that has more
best responses than the size of its support; in terms of
labels, this defines a point of~$P$ with more than $m$
labels or a point of $Q$ with more than $n$ labels.
An example is the degenerate game in Fig.~\ref{fthreat}
where the pure strategy $T$ (which is a mixed strategy
$x=(1,0)$ with a support of size one) has two best
responses, which is the point $((1,0),3)$ of $P$ in
Fig.~\ref{fupperlabel} with three labels $B,l,r$.
For extensive games, the strategic form is typically
degenerate, as the examples in Fig.\ \ref{f3}
and~\ref{fthreat} demonstrate.
In lrs, all arithmetic computations are done in {\em
arbitrary precision arithmetic} (rather than floating point
arithmetic with possible rounding errors) in order to
recognize with certainty that a basic variable is zero.
In addition, lrs uses the economical {\em integer pivoting}
method where basic feasible solutions are represented with
integers rather than rational numbers; see, for example, von
Stengel (2007, Section~3.5).

After enumerating all vertices of $P$ and $Q$ with lrs,
those pairs that fulfill the complementarity condition
(\ref{compl}) are the extreme equilibria of the game.
Avis et al.\ (2010) also describe an improved method
{\em lrsNash}, which is used in GTE, where only the vertices of
one polytope, say $P$, are generated, and the set $L$ of
labels {\em missing} from each such vertex $(x,v)$ is used
to identify the face $Q(L)$ of the other polytope that has
exactly the labels in~$L$.
If this face is not empty, then each of its vertices $(y,u)$
defines an extreme equilibrium $(x,y)$.
An alternative method to enumerate all extreme equilibria is
the EEE method of Audet et al.\ (2001) which performs a
depth-first search that chooses one tight inequality for
either player in each search step.
In Avis et al.\ (2010), EEE is implemented with exact
integer arithmetic and compared with {\em lrsNash}, and
performs better for larger games from size $15\times 15$
onwards.

Given the list of extreme equilibria, the set of all
equilibria is completely described as follows (for details
see Avis et al., 2010).
Consider the bipartite graph $R$ with edges $(x,y)$ that are
the extreme equilibria.
Each connected component $C$ of this graph defines a
topologically connected component of equilibria, as it is
output by GTE.
For each $C$, identify the maximal bipartite ``cliques'' in
$R$ of the form $U\times V\subseteq C$, that is, every pair
$(x,y)$ in $U\times V$ is an extreme equilibrium.
Then any convex combination of $U$ paired with any convex
combination of $V$ is also a Nash equilibrium, and every
equilibrium of the game can be represented in this way.
There may be several such cliques that define a component
$C$, as in Fig.\  \ref{fring} and~\ref{fKM} above.
In GTE, the maximal cliques $U\times V$ are computed with
an extension of the fast clique enumeration algorithm by 
Bron and Kerbosch (1973).


For a two-player game in strategic form, GTE computes
by default all Nash equilibria, as just described.
However, the state-of-the-art algorithms {\em lrsNash} and
EEE have exponential running time due the typically
exponential growth of the number of vertices of $P\times Q$.
These equilibrium enumeration algorithms therefore take
prohibitively long for games with more than 25, certainly 30
pure strategies for each player.
In addition, it is NP-hard to decide if a bimatrix game has
more than one Nash equilibrium (Gilboa and Zemel, 1989;
Conitzer and Sandholm, 2008), so one cannot expect to answer
that question quickly for larger games.

In practice, a finite game may represent a discretization of
an infinite game, where it may be useful to know its
equilibria when each player has several hundred strategies.
In that case, it is still possible to use methods that find
one equilibrium, such as the classic algorithm by Lemke and
Howson (1964).
This is a path-following algorithm that follows a sequence of
pivoting steps, similar to the simplex algorithm for linear
programming (for expositions see Shapley, 1974, or von
Stengel, 2002; 2007).
Like the simplex algorithm, it can take exponentially many
steps on certain worst-case instances (Savani and von
Stengel, 2006).
However, these seem to be rare in practice, and the
algorithm is typically fast for random games.

The Lemke--Howson algorithm has a free parameter, which is
a pure strategy of a player that together with the best
response to that strategy defines a ``starting point'' of
the algorithm in the space of mixed strategies $X\times Y$.
Varying this starting point over all pure strategies may
lead to different equilibria (however, if that is not the
case, the game may still have equilibria that are elusive to
the algorithm).

GTE uses a related but different method that is implemented
using the algorithm by Lemke (1965), which allows for
starting points that can be arbitrary mixed strategy pairs
$(\overline x,\overline y)$
(von Stengel, van den Elzen, and Talman, 2002).
In addition, the computation can be interpreted as the
``tracing procedure'' by Harsanyi and Selten (1988), where
players' strategies are best responses to a
weighted combination of their {\em prior}
$(\overline x,\overline y)$ and the actual strategies that
they are playing.
Initially, the players only react to the prior, whose weight
is then decreased; equilibrium is reached when the weight of
the prior becomes zero.
If starting from sufficiently many random priors gives always
the same equilibrium, this does not prove that this
equilibrium is unique, but may be considered as sufficient
reason to suggest this as a plausible way to play the game.

The computational experiments of von Stengel, van den Elzen,
and Talman (2002) show that for random games, the method
finds an equilibrium in a small number of pivoting steps,
and that typically many equilibria are found by varying the
prior by choosing it uniformly at random from the
strategy
simplices.
For a systematic investigation of this method in GTE, it
would be useful to generate larger games as discretizations
of games defined by arithmetic formulas for payoff
functions; this is a future programming project where a
closer integration with Gambit is envisaged.

\section{Equilibrium computation for games in extensive form} 
\label{s-sequenceform}

The standard approach to finding equilibria of an extensive
game is to convert it to strategic form, and to apply the
corresponding algorithms.
However, the number of pure strategies grows typically {\em
exponentially} in the size of the game tree.
In contrast, the {\em sequence form} is a strategic
description that has the same size as the game tree, which
is used in GTE.
We sketch the main ideas here; for more details see von
Stengel (1996; 2002) or von Stengel, van den Elzen, and
Talman (2002).

\begin{figure}[hbt]
\strut \hfill
\includegraphics[width=.75\hsize]{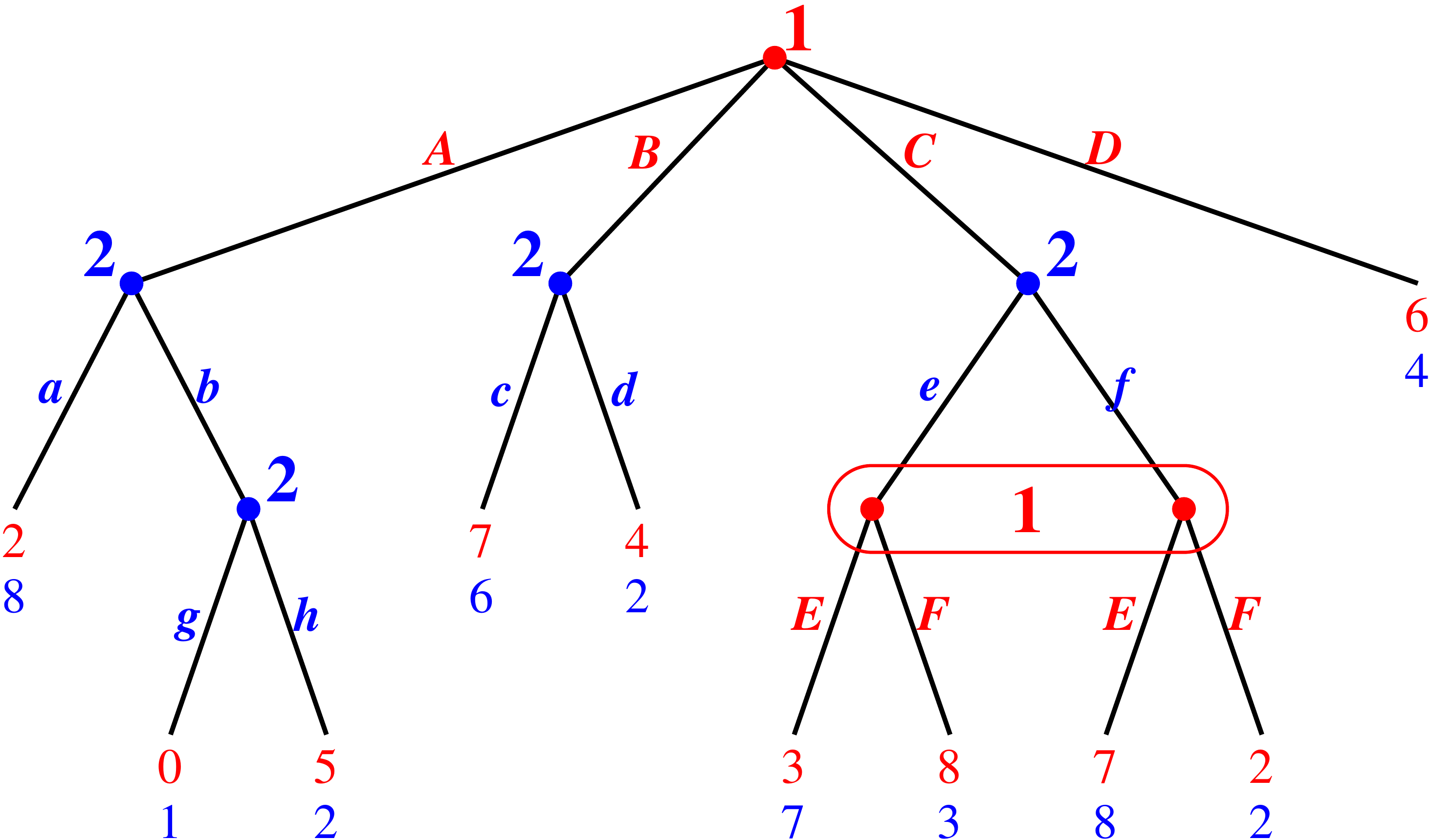} 
\hfill\strut
\caption{Extensive game with perfect information for
player~2 and no information for player~1.
}
\label{fef5}
\end{figure}

\begin{figure}[hbt]
\strut \hfill
\includegraphics[width=.75\hsize]{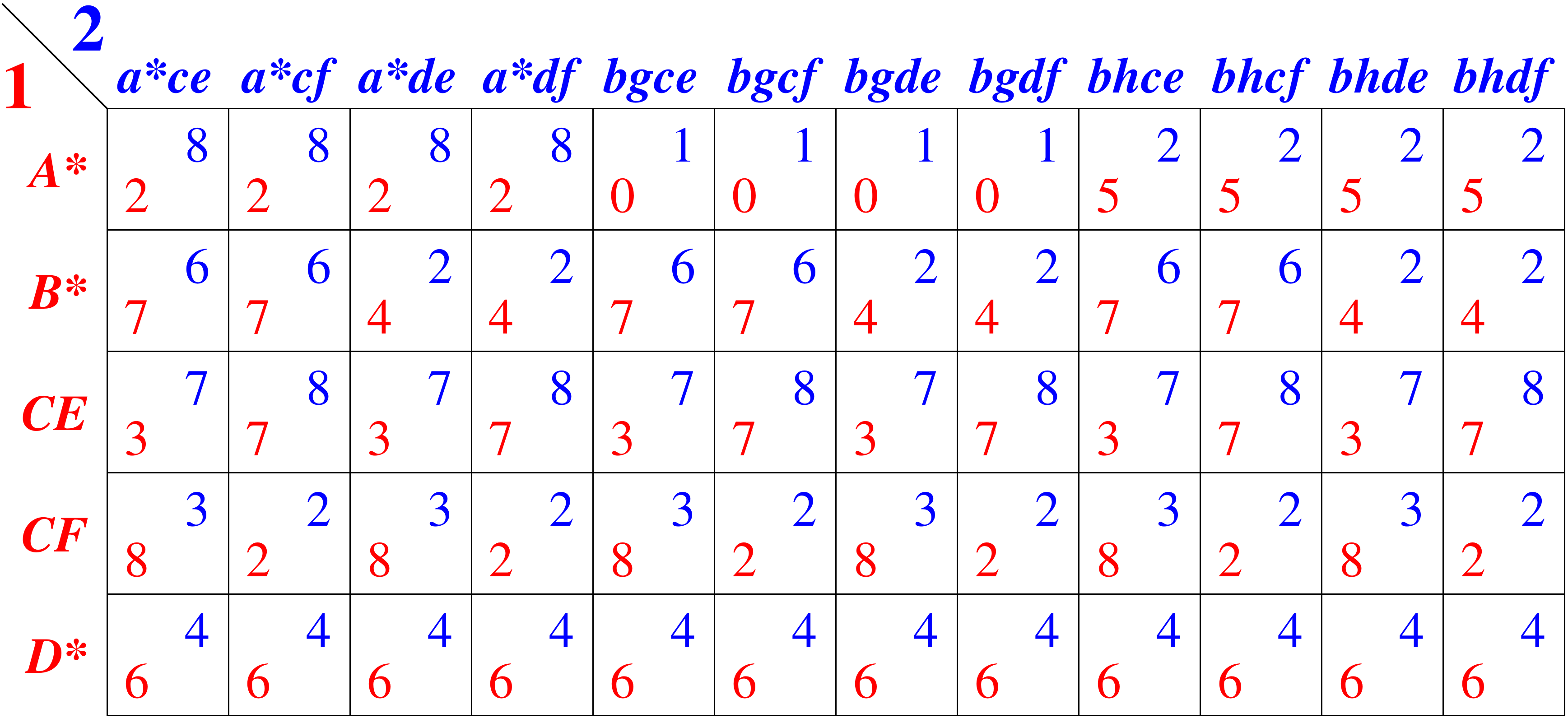} 
\hfill\strut
\caption{Reduced strategic form of the game in Fig.~\ref{fef5}.
In a reduced strategy, any move at an information set that
cannot be reached due to an earlier own move is left
unspecified and replaced by~``\,\7\,''.}
\label{fsf5}
\end{figure}

Fig.~\ref{fef5} shows an extensive game where player~1 has
four moves at the root and two moves at his second
information set.
Player~2 has perfect information and four singleton
information sets with two moves each.
A pure strategy of a player defines a move for every
information set, so player~1 has 8 and player~2 has 16 pure
strategies.  
Some of these pure strategies can be identified because they
describe the same ``plan'' of playing the game.
A {\em reduced} pure strategy does not specify a move at an
information set that is unreached due to an earlier own move
of the player.
For example, after move $a$ player~2 does not need to
specify move $g$ or~$h$ because the information set with
that move cannot be reached.
We represent such an unspecified move by an asterisk
``\,\7\,'' at the place where the move would be listed, so
$a\7ce$ is a reduced strategy where $\7$ stands for either
$g$ or~$h$, as in the strategy $bgce$ where no reduction is
possible.
The reduced strategic form is shown in
Fig.~\ref{fsf5}, which is a $5\times 12$ game.
Each cell shows the payoffs that result when the two
strategies meet.

In this game, a {\em mixed strategy} requires four
independent probabilities for player~1 (the fifth is then
determined because probabilities sum to one) and eleven for
player~2.
In general, the number of reduced strategies is {\em
exponential} in the size of the game tree, so there is a
large number of mixed strategy probabilities to compute.

\begin{figure}[hbt]
\strut \hfill
\includegraphics[width=.52\hsize]{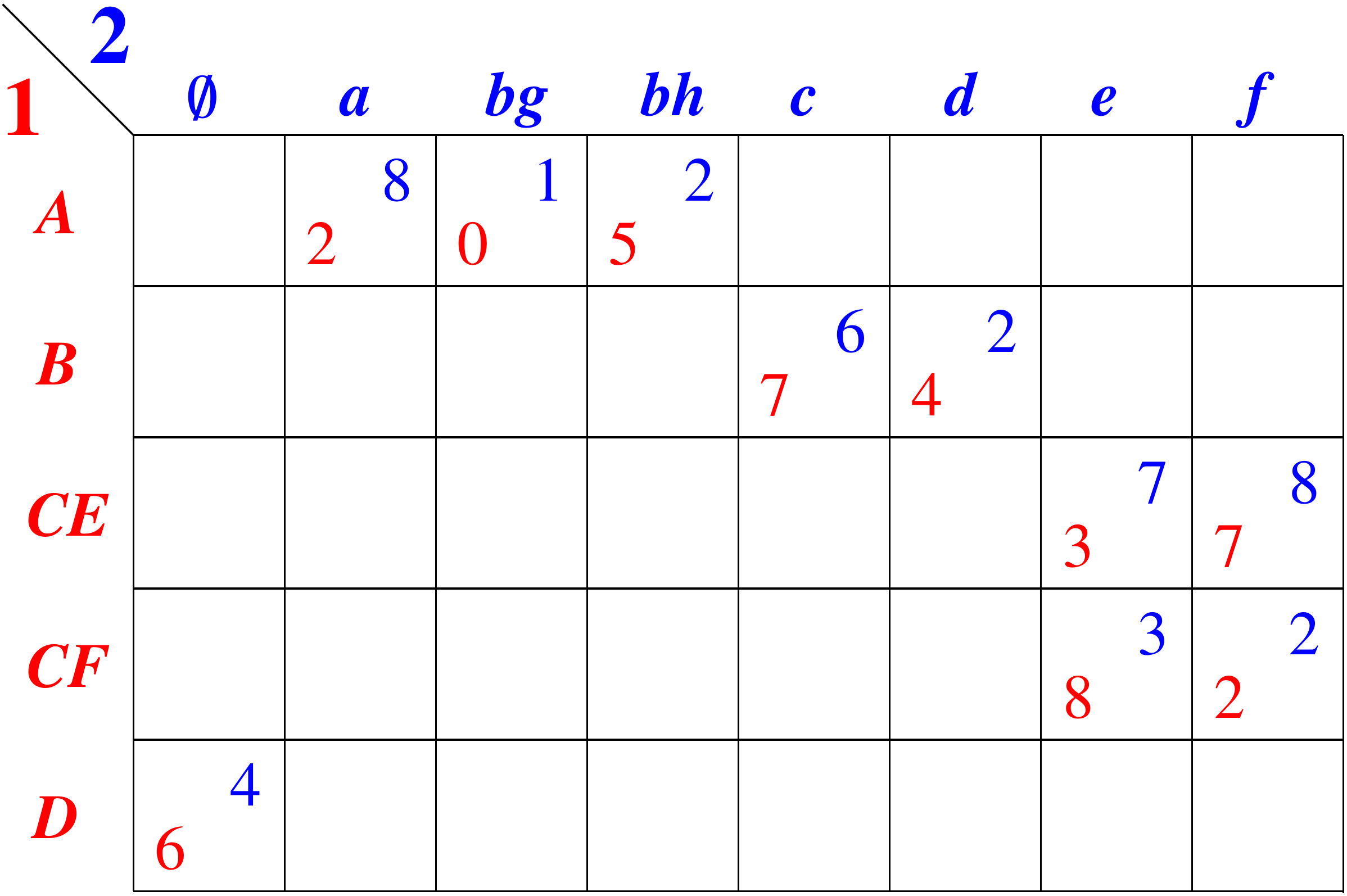} 
\hfill\strut
\caption{Sequence form payoffs of the game in Fig.~\ref{fef5}.
All empty cells have zero entries.
The rows are played with a probability distribution, whereas the
columns are played with weights $y_{\sigma}$ subject to the
equations
$y_{\emptyset}=1=y_a+y_{bg}+y_{bh}=y_c+y_d=y_e+y_f$.}
\label{fseq5}
\end{figure}

The {\em sequence form} is a compact strategic description
of the game that has the {\em same} size as the game tree.
Instead of strategies, it uses {\it sequences} of moves of a
given player along a path from the root to a leaf.
The payoff matrices are sparse and contain payoffs for those
pairs of sequences of the two players that lead to a leaf,
as shown in Fig.~\ref{fseq5} for the game in Fig.~\ref{fef5}.
This table is evidently more compact, and vastly more so for
larger games.
In this game, sequences and reduced strategies coincide for
player~1, because that player does not get any information
about the moves of the other player.
For player~2, her sequences $\sigma$ are played with
probabilities $y_{\sigma}$ that are not distributions over
the set of all sequences, but are subject to three separate
equations $y_a+y_{bg}+y_{bh}=1$, $y_c+y_d=1$, $y_e+y_f=1$.
These equations express the revealed information to player~2
following the respective moves $A$, $B$, and $C$ of
player~1, and can be derived systematically from the
structure of the information sets (with one equation per
information set, where we have substituted some equations
such as $y_b=y_{bg}+y_{bh}$ for the unused nonterminal
sequence~$b$; also, player~2's empty sequence $\emptyset$
leads here to the rightmost leaf and has constant
probability $y_{\emptyset}=1$).

The twelve reduced strategies of player~2 are in effect the 
combinations of one of the sequences $a,bg,bh$ combined with
one of $c,d$ and one of $e,f$, respectively.
In the sequence form, these sequences are randomized {\em
independently}.
This randomization translates to a {\em behavior strategy},
where the player randomizes locally over his moves at
each information set rather than globally over his pure
strategies as in a mixed strategy.
The underlying assumption about the information sets is that
of {\em perfect recall} which says that a player does not
forget what he knew or did earlier.
Using the sequence form implies the theorem of Kuhn (1953)
that a player with perfect recall can replace a mixed strategy 
by an equivalent behavior strategy.

In GTE, the sequence form is implemented with the
path-following algorithm of Lemke (1965) mentioned at the
end of the previous section, as described by von Stengel,
van den Elzen, and Talman (2002), which starts from an
arbitrary ``prior'' as a starting vector and finds one
equilibrium.
It can be applied to relatively large extensive games where
the strategic form is hopelessly large to be used.
For smaller extensive games, an enumeration of all Nash
equilibria based on the sequence form is implemented
as described by Huang (2011, Chapter~2); for a related
approach see Audet, Belhaiza, and Hansen (2009).
The number of independent probabilities in the sequence form
is at most that number in the reduced strategic form.
For example, player~2 in Fig.~\ref{fseq5} has four such
independent probabilities, like $y_a,y_{bg},y_c,y_e$,
as opposed to eleven for her reduced strategies; player~1 has
four independent probabilities for both sequences and
reduced strategies.
Huang's approach only uses these independent probabilities
and thus keeps a low dimension of the suitably defined
best-response polyhedra, which is important for the employed
vertex enumeration algorithms that have exponential running
time.


\section{Software architecture and development}
\label{s-software}

In this final section, we describe the architecture of the
GTE software on the server and client computers, and its
development.
The GTE program is open-source and part of the Gambit project,
with the goal of further integration with the existing
Gambit modules.
The development is time intensive and relies on the efforts
of volunteers, where we strongly invite any interested
programmer to contribute.


The GTE software is accessed by a web browser.
Behind its web address runs a {\em server} computer, which
delivers a webpage and the game-drawing software of GTE to
the user's {\em client} computer.
The user can then create games, and store them on
his client computer.
The {\em game solvers} for equilibrium computation
reside on the server.
They are invoked by sending a description of the game from
the client to the server over the internet.
When the algorithm has terminated, the server sends the text
with the computed equilibria back to the client where they
are displayed.

The client program with the graphical user interface (GUI)
is written in a variant of the JavaScript programming
language called ActionScript which is displayed with the
``Flash Player''.
Flash is a common software that runs as a ``plug-in'' on
most web browsers.
We chose Flash for GTE in 2010 because of the predictability
and speed of its graphical display.
However, Flash does not work on the iPad and iPhone mobile
devices (it does work on regular Apple computers), with the
more universally accepted HTML5 standard as its suggested
replacement.
In the future, we plan to replace the ActionScript programs
by regular JavaScript along with streamlining and further
improving the GUI part of GTE.

For security reasons, the client program requires active
permission of the user when writing or reading files on the
client computer.
When storing or loading a game as a local file, this is
anyhow initiated by the user and therefore causes no
additional delay.
We have designed a special XML format for games in extensive
or strategic form, where XML stands for ``extensible markup
language'' similar to the HTML language for web pages.
The tree is described by its logical structure.
For example, the leftmost leaf in Fig.~\ref{fthreat}
reached by move $T$ is encoded as follows:
\vskip-.5em

{\small
\begin{verbatim}
     <outcome move="T">
        <payoff player="1">1</payoff>
        <payoff player="2">3</payoff>
     </outcome>
\end{verbatim}}

\noindent
In addition, the file contains parameters for graphical
settings such as the orientation of the game tree.
The files can be converted to and from the file format used
by Gambit.
The XML description is also used to encode the game for
sending it to the server for running a game solver.

Communication with the server requires identifying the user
when there are different simultaneous client sessions, and
the use of the relevant internet protocols.
These are standard methods used for typical internet
transactions, which we use as freely available routines in
the Java programming language in which we have written the
server programs.

The first version of GTE was written on a publically
provided server (the Google App engine).
For security reasons, all programs on that platform have to
be written in Java, and are restricted in their ability to
access programs written in other programming languages,
called ``native'' code.
In particular, we could not use the C program lrs for vertex
enumeration, which was therefore rewritten in Java.
However, this requires to duplicate in Java any improvement
of lrs such as the {\em lrsNash} program mentioned in
Section~\ref{s-algostrat}, which is inefficient and prone to
errors.
In addition, it is difficult to incorporate other game
solvers.
For that reason, the web server for GTE is now provided on
a university computer, and uses the original lrs code of Avis
(2006).

GTE allows to compute all Nash equilibria of a two-player
game, but the running time increases exponentially with the
size of the game.
This means that there is a relatively narrow range of input
size (currently about 15 to 20 strategies per player) where
the efficiency of the implementation (such as a factor of
ten in running time) matters, and beyond which the
computation takes an impractically long time.
For small games, the game solvers would run satisfactorily
on the client using JavaScript, although at present there
are no good arbitrary precision arithmetic routines for this
programming language, apart from having to implement the
algorithms separately.
For games of ``medium'' size (below the ``exponential
barrier''), it is best to use a state-of-the-art algorithm
on the server, which works equally well for small games. 

However, one problem is that computations of, say, one hour
also use the computational resources of the server, which
degrade its performance, and which will be slow if there are
multiple such computations at the same time.
Here, we provide the option of installing the server on
any local computer, which then also works independently of
an internet connection.
This requires downloading and compiling the GTE code,
together with the necessary components of a Java compiler, a
lightweight server program called ``Jetty'' to communicate
locally with the web browser, and the (free) software for 
compiling the ActionScript part of GTE into an executable
Flash program.
This requires time and patience, probably more so than an
installation of Gambit, but offers the same interface as
that known from the web, so that the user may find the
investment worth it.

All GTE software is open source and free to use and alter
under the GNU General Public Licence (which requires derived
software to be free as well).
The software repository is at
\url{https://github.com/gambitproject/gte/wiki}
(including documentation), as part of the Gambit project
which has long been open-source.
The version control system and collaborative tool ``git''
stores the current and all previous versions of the
project.
Git efficiently stores snapshots of the entire project with
all modules and documentation, which can be branched off and
re-merged in decentralized development branches.
The {\tt github} repository for storing projects developed
with git is free to use for open-source software, and offers
a convenient web interface for administration and
documentation.

Open-source software is the appropriate way to develop an
academic project such as GTE which has limited commercial
use.
Academic software should be publically available for making
computations reproducible, which is increasingly recognized
as a need for validating scientific results that rely on
computation (Barnes, 2010).

On the algorithm side, one of the next steps is to integrate
other existing game solvers.
Foremost among these are the algorithms already implemented
in Gambit, which should be easily incorporated via their
existing file formats for input and output.
These include (see McKelvey, McLennan, and Turocy, 2010)
algorithms for finding Nash equilibria of more than two
players with polynomial systems of equations, or
via iterated polymatrix approximation (Govindan and Wilson,
2004),
or simplicial subdivision (van der Laan, Talman,
and van der Heyden, 1987), 
and the recent implementation of ``action-graph games''
(Jiang, Leyton-Brown, and Bhat, 2011).
Moreover, larger games should not be created manually but 
automatically.
One focus of the current development of Gambit is to provide
better facilities for the automatic generation of games
using the Python programming language.

So far, the focus of GTE development has been to design a
robust and attractive graphical ``front end'' for creating
and manipulating games.
Such an implementation does not count as scholarly research,
so it is difficult to fund it directly with a research grant,
or to let a PhD student invest a lot of time in it. 
As a topic for an MSc thesis or undergraduate programming
project, usually too much time is needed to become familiar
with the existing code or the necessary background in game
theory.
The Gambit project and GTE were sponsored by studentships
from the ``Google Summer of Code'' in 2011 and 2012.
Here, the most successful projects for GTE were those where
the students could bring in their technical expertise on web
graphics, for example, to complement our background in
game-theoretic algorithms.

As with open-source software in general, the development of
GTE relies on the efforts, competence, and enthusiasm of
volunteers, where we were lucky to get help from excellent
students listed in the Acknowledgments.
We invite and encourage any interested programmer to
contribute to this project.

\vskip3ex\noindent\footnotesize\textbf{Acknowledgments }
We are indebted to Mark Egesdal, Alfonso G\'omez-Jordana and
Martin Prause for their invaluable contributions as
programmers of GTE.
Mark Egesdal designed the main program structure and coined
the name ``Game Theory Explorer'', with financial support
from a STICERD research grant at the London School of
Economics in 2010.
Alfonso G\'omez-Jordana and Martin Prause were funded by the
Google Summer of Code in 2011 and 2012 for the open-source
Gambit project, and continue as contributing volunteers.
Karen Bletzer wrote conversion programs between GTE and Gambit
file formats.
Wan Huang implemented the enumeration of Nash equilibria
based on the sequence form.
We thank Theodore L. Turocy for inspiring discussions and
for his support of GTE as part of Gambit.  
David Avis has written the {\em lrsNash} code for 
computing all Nash equilibria of a two-player game.
All contributions and financial support are gratefully
acknowledged.



\section*{References}
\addcontentsline{toc}{section}{References} 
\frenchspacing
\parindent=-2em\advance\leftskip by2em
\parskip=.25ex plus.1ex minus.1ex
\small\footnotesize
\hskip\parindent
Audet, C., S. Belhaiza, and P. Hansen (2009),
A new sequence form approach for the enumeration of all
extreme Nash equilibria for extensive form games.
International Game Theory Review 11, 437--451.

Audet, C., P. Hansen, B. Jaumard, and G. Savard (2001),
Enumeration of all extreme equilibria of bimatrix games.
SIAM Journal on Scientific Computing 23, 323--338.

Avis, D. (2000),
lrs: a revised implementation of the reverse search
vertex enumeration algorithm. In: Polytopes--Combinatorics and Computation,
eds. G. Kalai and G. Ziegler, DMV Seminar Band 29,
Birkh\"auser, Basel, pp. 177--198.

Avis, D. (2006),
User's Guide for lrs.
\url{http://cgm.cs.mcgill.ca/~avis}.

Avis, D., G. Rosenberg, R. Savani, and B. von Stengel (2010),
Enumeration of Nash equilibria for two-player games.
Economic Theory 42, 9--37.

Bagwell, K. (1995),
Commitment and observability in games.
Games and Economic Behavior 8, 271--280.

Balthasar, A. V. (2009),
Geometry and Equilibria in Bimatrix Games.
PhD Thesis, London School of Economics.

Barnes, N. (2010),
Publish your computer code: it is good enough.
Nature 467, 753--753.

Belhaiza, S. J., A. D. Mve, and C. Audet (2010),
XGame-Solver Software.
\url{http://faculty.kfupm.edu.sa/MATH/slimb/XGame-Solver-Webpage/index.htm} 

Bron, C., and J. Kerbosch (1973),
Finding all cliques of an undirected graph.
Communications of the ACM 16, 575--577.


Conitzer, V., and T. Sandholm (2008),
New complexity results about Nash equilibria.
Games and Economic Behavior 63, 621--641.  



Datta, R. S. (2010),
Finding all Nash equilibria of a finite game using
polynomial algebra. Economic Theory 42, 55--96. 

Gilboa, I., and E. Zemel (1989),
Nash and correlated equilibria: some complexity
considerations.
Games and Economic Behavior 1, 80--93.


Govindan, S., and R. Wilson (2004),
Computing Nash equilibria by iterated polymatrix approximation.
Journal of Economic Dynamics and Control 28, 1229--1241.

Harsanyi, J. C., and R. Selten (1988),
A General Theory of Equilibrium Selection in Games.
MIT Press, Cambridge, MA.  

Hauk, E., and S. Hurkens (2002),
On forward induction and evolutionary and strategic
stability.
Journal of Economic Theory 106, 66--90.

Huang, W. (2011),
Equilibrium Computation for Extensive Games.
PhD Thesis, London School of Economics.

Jiang, A. X., K. Leyton-Brown, and N. A. R. Bhat (2011),
Action-graph games.
Games and Economic Behavior 71, 141--173.

Kohlberg, E., and J.-F. Mertens (1986),
On the strategic stability of equilibria.
Econometrica 54, 1003--1037.

Kuhn, H. W. (1953),
Extensive games and the problem of information.
In: Contributions to the Theory of Games II, eds.
H.~W. Kuhn and A.~W. Tucker, Annals of Mathematics Studies~28,
Princeton Univ. Press, Princeton, pp.\ 193--216.  

Langlois, J.-P. (2006),
GamePlan, a Windows application for representing and solving
games. 
\url{http://userwww.sfsu.edu/langlois/}

Lemke, C. E. (1965),
Bimatrix equilibrium points and mathematical
programming. Management Science 11, 681--689.

Lemke, C. E., and J. T. Howson, Jr. (1964),
Equilibrium points of bimatrix games.
Journal of the Society for Industrial and
Applied Mathematics 12, 413--423.  

McKelvey, R. D., A. M. McLennan, and T. L. Turocy
(2010), Gambit: Software Tools for Game
Theory, Version 0.2010.09.01.
\url{http://www.gambit-project.org}

Nash, J. (1951),
Noncooperative games. Annals of Mathematics 54, 286--295.

Osborne, M. J. (2004),
An Introduction to Game Theory.
Oxford University Press, Oxford.


Quint, T., and M. Shubik (1997),
A theorem on the number of Nash equilibria in a bimatrix
game.
International Journal of Game Theory 26, 353--359.

Savani, R. (2005),
Solve a bimatrix game.
Interactive website at
\url{http://banach.lse.ac.uk/}

Savani, R., and B. von Stengel (2004),
Exponentially Many Steps for Finding a Nash Equilibrium in a
Bimatrix Game.
CDAM Research Report LSE-CDAM-2004-03. 

Savani, R., and B. von Stengel (2006),
Hard-to-solve bimatrix games. Econometrica 74, 397-429. 

Shapley, L. S. (1974),
A note on the Lemke--Howson algorithm.
Mathematical Programming Study 1: Pivoting and Extensions,
175--189.  

van der Laan, G., A. J. J. Talman, and L. van der Heyden
(1987),
Simplicial variable dimension algorithms for solving the
nonlinear complementarity problem on a product of unit
simplices using a general labelling.
Mathematics of Operations Research 12, 377--397.

von Stengel, B. (1996),
Efficient computation of behavior strategies.
Games and Economic Behavior 14, 220--246.

von~Stengel, B. (1999),
New maximal numbers of equilibria in bimatrix games.
Discrete and Computational Geometry 21, 557--568.

von~Stengel, B. (2002),
Computing equilibria for two-person games.
In: Handbook of Game Theory, Vol.~3, eds. R. J. Aumann and
S. Hart, North-Holland, Amsterdam, pp. 1723--1759.

von~Stengel, B. (2007),
Equilibrium computation for two-player games in strategic
and extensive form.
In: Algorithmic Game Theory, eds. N. Nisan et al.,
Cambridge Univ. Press, Cambridge, pp. 53--78. 

von~Stengel, B. (2012),
Rank-1 games with exponentially many Nash equilibria.
arXiv:1211.2405. 

von~Stengel, B., A. H. van den Elzen, and A. J. J. Talman
(2002),
Computing normal form perfect equilibria for extensive
two-person games.
Econometrica 70, 693--715.


\end{document}